\documentclass[a4paper]{aa}
\usepackage{epsfig,times}

\begin{document}
\def\ni{\noindent} 
\def\ea{{ et\thinspace al.}}                        
\def\eg{{ e.g.}\ }                                    
\def\ie{{ i.e.}\ }                                   
\def\cf{{ cf.}\ } 
\def\rot{\mathop{\rm curl}\nolimits}
\def\div{\mathop{\rm div}\nolimits} 
\def\grad{\mathop{\rm grad}\nolimits} 
\def\solar{\ifmode_{\mathord\odot}\else$_{\mathord\odot}$\fi} 
\def\gsim{\lower.4ex\hbox{$\;\buildrel >\over{\scriptstyle\sim}\;$}} 
\def\lsim{\lower.4ex\hbox{$\;\buildrel <\over{\scriptstyle\sim}\;$}}  
\def\~  {$\sim$} 
\def\cl {\centerline} 
\def\rl {\rightline} 
\def\x	{\times} 
\def\alf{$\alpha$}
\def\L{$\Lambda$}
\def\Om{$\Omega$}
\def\nT{$\nu_{\rm T}$\ }
\def\mT{$\mu_{\rm T}$\ }
\def\cT{$\chi_{\rm T}$\ }
\def\eT{$\eta_{\rm T}$\ }
\def\Omst{$\Omega^*$}
\def\apj{{ApJ}\ }       
\def\apjs{{ Ap. J. Suppl.}\ } 
\def\apjl{{ Ap. J. Letters}\ } 
\def\pasp{{ Pub. A.S.P.}\ } 
\def\mn{{MNRAS}\ } 
\def\aa{{A\&A}\ } 
\def\aasup{{ Astr. Ap. Suppl.}\ } 
\def\baas{{ Bull. A.A.S.}\ } 
\def\csss{{Cool Stars, Stellar Systems, and the Sun}\ }
\def\an{{Astron. Nachr.}\ }
\def\sp{{Solar Phys.}\ }   
\def\gafd{{Geophys. Astrophys. Fluid Dyn.}\ } 
\def\ass{{Ap\&SS}\ }
\def\acta{{Acta Astron.}\ }
\def\jfm{{J. Fluid Mech.}\ }
\def\AIP{Astrophysikalisches Institut Potsdam}
\def\F{Ferri\`{e}re}
\def\R{R\"udiger}
\def\E{Elstner}
\def\K{Kitchatinov}
\def\qq{\qquad\qquad}                      
\def\qqq{\qquad\qquad\qquad}               
\def\q{\qquad}
\def\DR{differential rotation\ }
\def\bib{\item{}}
\def\top{\item}
\def\toptop{\itemitem}
\def\start{\begin{itemize}}
\def\stop{\end{itemize}}
\def\beg{\begin{equation}}
\def\ende{\end{equation}}
\def\la{\langle}
\def\ra{\rangle}


\title{Global accretion disk simulations of magneto-rotational instability}   
\author{R. Arlt \and G. \R}
\offprints{G. \R}
\institute{Astrophysikalisches Institut Potsdam,  An der Sternwarte 16, 
D-14482 Potsdam, Germany}
\thesaurus{(02.01.2,02.09.1,02.13.1,02.13.2,02.20.1)}
\date{\today}
\maketitle
\markboth{R. Arlt \& G. \R: Global simulations of accretion disks} 
{R. Arlt \& G. \R: Global simulations of accretion disks}

\begin{abstract}
We perform global three-dimensional simulations 
of accretion disks integrating
the compressible, non-viscous, but diffusive MHD equations.
The disk is supposed to be isothermal. We make use of the
ZEUS-3D code integrating the MHD equations and added magnetic
diffusivity.
We measure the efficiency of the angular-momentum transport.
Various model simulations delivered transport parameters of
$\alpha_{\rm SS}=0.01$ to 0.05 which are consistent with 
several local numerical investigations. Two of the models
reach a highly turbulent state at which $\alpha_{\rm SS}$ is
of the order of 0.1. After a certain stage of saturating of the
turbulence, Reynolds stress is found to be negative (inward transport)
in many of the models, whereas Maxwell stresses dominate and deliver a 
positive (outward) total transport. Several of the models
yield strongly fluctuating Reynolds stresses,
while Maxwell stresses are smooth and always transport outwards.
Dynamo action is found in the accretion disk simulations.
A positive dynamo-$\alpha$ is indicated in the northern
hemisphere of the most prominent run, coming along with 
negative kinetic and current
helicities (all having the opposite sign on the southern
side). The dipolar structure of the magnetic field is
maintained throughout the simulations, although indication
for a decay of antisymmetry is found. The simulations covered
relatively thick disks, and results of thin-disk dynamo
models showing quadrupolar fields may not be compatible with
the results presented here.

\end{abstract}
\section{Introduction}
Accretion processes in astrophysical disks lead to enormous
luminosity and huge changes in disk structure during astronomically
short times. Efficient transport mechanisms are necessary to
achieve such short time-scales. Anisotropic turbulence appears
to be a major physical condition to provide astrophysical disks
with strong transport. As these disks generally exhibit increasing
specific angular momentum towards larger radii and thus fulfill the
stability criterion of Rayleigh, they do not give rise to an instability
leading to turbulence by themselves.
Searches for instabilities in disks with rotation profiles
similar to a Keplerian one unveiled several ways to turbulence
being more or less favorable with respect to their
prerequisites for the disk configuration. Gravitational instability
needs the disk to be either cool or massive. Nonlinear and
nonaxisymmetric perturbations require a severe additional perturber
near the disk; conditions for instability in a purely hydrodynamical
disk were derived by Dubrulle (1993). Hydrodynamic instability 
essentially comes down to violating the Rayleigh criterion saying 
that a rotation profile with an increasing specific angular momentum with 
radius is hydrodynamically stable, that is for 
$\partial l^2/\partial r>0$. Since the typical length of
perturbations caused be external forces is supposed to be very
large, the required amplitude of the perturbations has to
be considerable, too, in order to violate the Rayleigh criterion
locally. If the length scale of the perturbation is comparable
to the radius, a strong alteration of the Keplerian velocity of
several per cent is needed. Convection was shown to deliver 
either negligible transport (Stone \& Balbus 1996) or inward
angular momentum transport (Kley et~al. 1993). 

The requirements
for the magnetic shear-flow instability (Balbus \& Hawley 1991)
do match astrophysical conditions in accretion disks in many 
configurations. All it needs is a radially decreasing
angular velocity and a weak magnetic field threading the
rotating object. It can even be shown that the temperature
range applicable to the magnetic shear-flow 
concept is very broad; even very small ionization fractions are sufficient
to magnetize a disk in many cases (Balbus \& Hawley 1998).

First numerical approaches to the magnetic shear-flow instability
tackled the local problem; the linear analysis as described
by Balbus \& Hawley (1991) were immediately followed by
2D simulations (Hawley \& Balbus 1991) of a small box-shaped 
domain which was cut out of the disk. These computations
confirmed the relation between magnetic-field strength and
wavenumber derived from the linear analysis earlier, they 
showed that the maximum growth rate of any wavenumber is
independent from the field strength and that the system is capable of
transporting angular momentum. Because of being restricted
to axisymmetric problems, they could not provide self-sustained
turbulence which needs dynamo-action, and the slow decay of the 
turbulence is an indirect effect of the Cowling theorem. 
Improved computations dealt with
a three-dimensional even though local configuration, and particular
care was taken for the radial boundary conditions which are
not simply periodic, but account for the shear due to differential
rotation (Hawley et al.~1995). These comprehensive computations
indeed resulted in magnetically sustained turbulence whose anisotropy
causes efficient outward angular momentum transport. This work 
was followed up by stratified models (Stone et al. 1996) covering
more than 50~orbital periods of the box cut-out. As the computational
domain covered 2 density scale heights, this work was a first
step towards global simulations, followed up by similar approaches
such as Ziegler \& R\"udiger (2000a,\,b).

Linear studies of global configurations of disks threaded by 
magnetic fields in various directions were carried out. Curry \& Pudritz
(1995) investigated the stability for vertical and azimuthal
fields threading the disk. They found in detail that the actual
initial field geometry does not strongly depend on field topology
as was suggested by the numerical simulations of Hawley \& Balbus (1991).
R\"udiger et al. (1999) particularly
addressed the angular momentum transport in their linear study.
These investigations are now followed by nonlinear simulations
of a compressible fluid with density stratification in global 
computational domains. The most recent global approach was
presented by Hawley (2000) following the evolution of a thick torus
under the influence of an external magnetic field threading parts
of the computational domain. The magnetic shear-flow instability
was found to set in quickly causing enough turbulent viscosity
to soon form a Keplerian velocity profile. All the above mentioned
numerical studies integrate the ideal MHD equations omitting 
magnetic diffusivity. Local simulations including diffusivity were
performed by Fleming et al.\ (2000) which proved the onset of instability
even for low conductivity.

The full understanding of accretion disks implies self-excited
dynamo action as well as angular momentum transport. Can positive 
angular momentum transport be coupled with a suitable kinetic
helicity providing the expected dynamo action according to the 
$\alpha$-effect principle? As the kinetic helicity in stratified,
rotating disks is expected to be negative, a wrong sign (positive)
would follow for the dynamo-$\alpha$-effect according to the
conventional $\alpha$-theories. We present global three-dimensional 
diffusive simulations and study the angular momentum transport 
as well as dynamo action and the sign of $\alpha^{\rm dyn}$ as a
consequence of the correlation with the flow.

The present computations are not meant as particular simulations of a star
formation process. They focus on the applicability of the 
magnetic shear-flow instability for an astronomically fast
process like the formation of a star. Such global simulations are still 
a challenge for modern computers and fast algorithms.


\section{The simulation setup}
The computations presented here make use of the ZEUS-3D
code developed for astrophysical problems of magnetohydrodynamics
(see the key papers Stone \& Norman 1992a,b; Stone et al. 1992 for
numerical, Clarke et al.\ 1994 for technical details). We use cylindrical
coordinates and an extensive computational domain covering 
radii $r=5$ to 6 and $r=4$ to 6, 
a vertical extension of $h=-1$ to +1,
and the full azimuthal range of $\phi=0$ to $2\pi$. See 
Table~\ref{tab1} for a list of models presented in this Paper.
In this approach, we assume an isothermal disk to save computation
time on the energy equation. The remaining system for
integration is
\begin{eqnarray}
\frac{\partial\rho}{\partial t}+\div(\rho {\vec u})&=&0\\
\frac{\partial{\rho \vec u}}{\partial t}+\div(\rho{\vec u}{\vec u})
&=&-\grad p-\rho \grad\Phi+{\vec J}\times {\vec B}\\
\frac{\partial {\vec B}}{\partial t}&=&\rot({\vec u} \times {\vec B})
+\eta\triangle \vec B,
\end{eqnarray}
where $\rho$, ${\vec u}$, and ${\vec B}$ are the density, velocity, and magnetic
field resp.; $p$ is the pressure ($p\propto \rho$ in our model),
$\Phi$ is the gravitational potential (solely from a central 
mass $M$), ${\vec J}$ is the current density,
and $\eta$ is the magnetic diffusivity which is not an original
ingredient to the ZEUS code. It is constant in time and space.
The computations of the electromotive force in the routines
{\tt emfs}, {\tt mocemfs}, and {\tt hsmoc} of ZEUS-3D are extended with the
appropriate $\eta\rot\vec B$ components. We chose $\eta=0.001$ to 0.01.
The additional time-step criterion resulting from the diffusivity
is roughly 0.1 for the finest grid used here. It is therefore
irrelevant for the determination of the time-step which is typically
$10^{-4}$ or, in the case of strong fluctuations of velocity
and magnetic field, one or two orders of magnitude lower. In
physical units, this diffusivity is large and in fact accounting
for a subgrid turbulent diffusivity which cannot be resolved.

\begin{table*}
\caption{\label{tab1}Model parameter used for this study. $T_0$ is the
time when the magnetic field was switched on (in orbital units). The
$z$-range is always from $-1.0$ to $+1.0$. If accretion boundary conditions
occur, they refer to the outer boundary; the inner boundary is `outflow'
then.}
\begin{small}
\begin{tabular}{cllcccl}
\hline
Model&Resolution ($z$, $r$, $\phi$)&Radial boundary condition&$r$-range&$T_{\rm orb}$ &$T_0$& $\eta$\\
\hline
Ia    & $31\times61\times351$ & inner outflow & 5.0--6.0&0.222 & \phantom{0}4.5  &0.001\\
Ib    & $31\times61\times351$ & all outflow& 5.0--6.0&0.222 & \phantom{0}4.5 &0.001\\
II    & $31\times61\times351$ & accretion $u_{\rm in}=-0.001c_{\rm ac}$& 4.0--6.0&0.159 & \phantom{0}9.4 &0.001\\
III   & $31\times61\times351$ & accretion $u_{\rm in}=-0.01c_{\rm ac}$& 4.0--6.0&0.159 & 41.2  &0.001 \\
IV   & $31\times31\times351$ & accretion $u_{\rm in}=-0.001c_{\rm ac}$& 5.0--6.0&0.222 & 26.8 &0.001\\
V    & $31\times61\times351$ & Gaussian accretion $u_{\rm in}=-0.001c_{\rm ac}$& 4.0--6.0& 0.159 & 29.6  &0.01\\
VI   & $31\times61\times351$ & Gaussian accretion $u_{\rm in}=-0.01c_{\rm ac}$& 4.0--6.0&  0.159 & 18.9 & 0.01\\
VII & $31\times31\times151$ & Gaussian accretion $u_{\rm in}=-0.001c_{\rm ac}$&
 4.0--6.0& 0.159 & 16.1 & 0.01\\
VIII& $31\times61\times351$ & accretion $u_{\rm in}=-0.001c_{\rm ac}$& 3.0--7.0
& 0.103 & 24.7 & 0.001\\
\hline
\end{tabular}
\end{small}
\end{table*}

The gravitational potential is spherically symmetric, whence
the $z$-component of the gravitation is retained within the disk.
We therefore obtain a density stratification unlike the computations
by Armitage (1998) who omits the $z$-component of the gravitation
and applies periodic boundary conditions for the upper and lower
boundaries. The sound speed is always $c_{\rm ac}=10$ which is
roughly $0.07u_{\rm K}$ in terms of the average Keplerian velocity
in the simulated ring. The initial density distribution is 
Gaussian with $r$-dependent density scale-height, $\rho(r,z)=
\rho_{\rm c}\exp(-z^2/H(r)^2)$.

The magnetic field threads the disk vertically with a mere
$z$-component. We tried a homogeneous initial field $B_z(r)={\rm const}$
for the $z$-boundaries as well as a field
\begin{equation}
  B_z(r,t=0) = b_0 r^{-1} \sin [2\pi (r-r_{\rm i})/(r_{\rm o}-r_{\rm i})]
\end{equation}
-- where $r_{\rm i}$ and $r_{\rm o}$ are the inner and outer boundary
radii resp. -- which has zero total magnetic flux through the $z={\rm const}$
surfaces. The choice of the initial field will have implications
for the topology of the field later on. As the magnetic flux through
the surfaces is kept constant, the first approach will always
preserve a non-zero average magnetic field through the vertical
boundaries, whereas the field penetrating the top and the bottom
of the computational domain can completely vanish in the second
choice of initial fields. The results presented here were obtained 
with the second approach of vanishing average initial field $B_z$.
The parameter $b_0$ was chosen between 50 and 100 giving an amplitude of the
initial magnetic field of ${\rm max}(B_z)=10$ to 20 corresponding to
maximum Alfv\'en speeds at the upper and lower boundaries of
$u_{\rm A}=13$ to 26 or $u_{\rm A}=1.3\,c_{\rm ac}$ to $2.6\,c_{\rm ac}$.
The Alfv\'en velocity in the equatorial plane is subthermal
with $u_{\rm A}=0.32$ to 0.63 or $0.032\,c_{\rm ac}$ to $0.063\,c_{\rm ac}$.

The initial velocity field is a merely Keplerian
motion following $u_\phi=\sqrt{GM/r}$, where $G$ is the gravitational
constant and $M$ is the central mass which is $10^5$ in our computational 
units. The total disk mass is 35,000 for Model Ia, Ib, and IV
and 56,000 for the other models.
Times are henceforth measured in
orbital periods which convert by $T_{\rm orb}=0.159$ from the arbitrary units
of the code.

The number of cells in each coordinate direction was 
$31\times 61\times 351$ for the $z$-, $r$-, and
$\phi$-directions. Tests with up to 621 cells in
$\phi$-direction have been carried out, but the increased
computation time does not allow reasonable periods to be covered
by the simulation. The aspect ratio of the grid cells
is not unity. Finest resolution is achieved in radial
direction, lowest in azimuthal direction:
$\Delta z/\Delta r=2$, $r\Delta\phi/\Delta r=2.18$ for the
smallest radius and $r\Delta\phi/\Delta r=3.28$ for the outer
radial edge. The fastest growing mode of the magneto-rotational
instability has a wavelength of $\lambda_{\rm inst}=
2\pi\sqrt{16/15}\,u_{\rm A}/\Omega$. With the angular velocity
of $\Omega=28.2$ in the middle of the computation domain,
we obtain wavelengths between 0.4 and 0.9 with the above
given Alfv\'en velocities of the initial field strength.
These wavelengths are upper limits as they refer to the
sites of maximum field and minimum density which need not
coincide necessarily. Yet, they are well resolved by the
computation domain covering $z$-ranges of 2.0 and two
different $r$-intervals of 1.0 and 2.0. 

The integration of the magnetic fields was done with the
evolution of the electromotive forces by the 
Constraint Transport scheme (CT, Evans \& Hawley 1988) 
which preserves a divergence-free configuration.

Two terms adding a source of viscosity to the hydrodynamic
equations extend the above Navier-Stokes equation. The 
von Neumann-Richtmyer artificial viscosity depends on the
square of velocity gradients and acts only on decreasing velocity
in the direction of propagation, i.e.\ compression. The strength
of this term is denoted by $q_{\rm con}$ in the implementation
of the ZEUS-3D code. The second
term depends linearly on the velocity gradient and acts on both
compression and expansion; the strength is represented by
$q_{\rm lin}$. Tests with a stable Keplerian flow found the
choice of $q_{\rm con}=0.1$ and $q_{\rm lin}=1.0$ to be suitable.
These values were used throughout all the runs described here.
The Courant number determining the 'safety' of a certain maximum
allowed time step derived from the velocities, magnetic fields, and
the artificial viscosities, is set to 0.5.

The ZEUS-3D code provides three methods of interpolation in
the transport step: the first-order donor-cell method, 
the second-order van-Leer method, and the third-order
piecewise-parabolic-advection method. The original
ZEUS-3D copy chooses the second-order interpolation for
the transport of all quantities, and we have not altered
this switch for the computations given here.

Occasionally, pockets of extremely low density may emerge
coupled with extraordinarily high speeds exceeding the
Keplerian velocity. We suppress the existence of such 
pockets by a mass replenishment as soon as the density
of a certain cell drops below $10^{-4}$ the central density.
The actual mass being thus added to the total disk mass
is found to be negligible. The mass replenishment thus
acts like a lower limit for the time step.



\subsection{Boundary conditions}
The vertical boundary condition on the $z={\rm const}$-faces
of our computational domain are closed for the flow, i.e.\ 
$v_z=0$. The boundary condition is stress-free in the sense
of $\partial v_r/\partial z=0$ and $\partial v_\phi/\partial z=0$.
The magnetic field has to fulfill the `opposite' boundary
condition as it is only allowed to cross the boundary
perpendicularly, which is actually implemented affecting the
electromotive force ${\vec{\cal E}}=\vec u\times \vec B - \eta \rot \vec B$
such that ${\cal E}_z=0$,  $\partial {\cal E}_r/\partial z=0$, and
$\partial {\cal E}_\phi/\partial z=0$.

No magnetic field lines are allowed to penetrate the radial boundary 
surfaces. Computations were carried out with two choices of radial
boundary conditions regarding the velocity. 
The first does not allow flow through the
top, bottom and outer radial boundaries. The inner boundary was chosen 
to be `outflow', that is, the physical arrays are constantly extrapolated
into the ghost zones of the computational domain. The only exception is
an inflow, in our case a $u_r>0$ at the inner boundary. This velocity
component is reset to zero then. All radial velocities $u_r<0$ are
copied to the ghost zones beyond the computational domain. We will
henceforth refer to these models as Model Ia (only inner boundary
has `outflow') and Ib (all both radial and both vertical boundaries
have `outflow').

Since the outflow condition is likely to empty the disk on the viscous
time-scale, the second choice tries to account for the accreted matter
and feeds the disk at the outer radial boundary. We sum up the mass
loss rate at the inner boundary $\dot M=r_{\rm min}\Delta\phi\Delta z
\sum_i\sum_k \rho_{ik} (u_{r})_{ik}$ for
the total mass loss. A constant, small inflow velocity $u_{\rm in}$
is assumed for the outer boundary. Accordingly, the density at the
outer boundary is determined to account for the mass flux at the
inner boundary. Since we use the average influx, its $z$-dependence
(nor even the $\phi$-dependence) is not used for a $z$- (nor even $\phi$-)
dependent inflow from far radii. We tested inflow conditions with
homogeneous and Gaussian and density distribution.
It was supposed that an average
influx of mass will establish according to the current viscosity
(either numerical or turbulent) in the disk. These models are refered
to as Model II-VIII in the following. 

The azimuthal velocity, $u_\phi$, is extrapolated into the radial boundary
zones by a power law $r^{-1.5}$ based on the last zone of the
computational domain. The velocities are thus not Keplerian,
they just follow the same radial power law.
Since we cover the full azimuthal range in $\phi$, the 
boundaries at the $\phi={\rm const}$ surfaces are periodic, naturally. 

Even though the ZEUS-3D code is remarkably stable for a broad
variety of configurations and physical problems, a comprehensive
study of the influence of the computational schemes provided
were found to be inevitable for trustworthy results.

\begin{figure}
  \begin{center}
    \epsfig{file=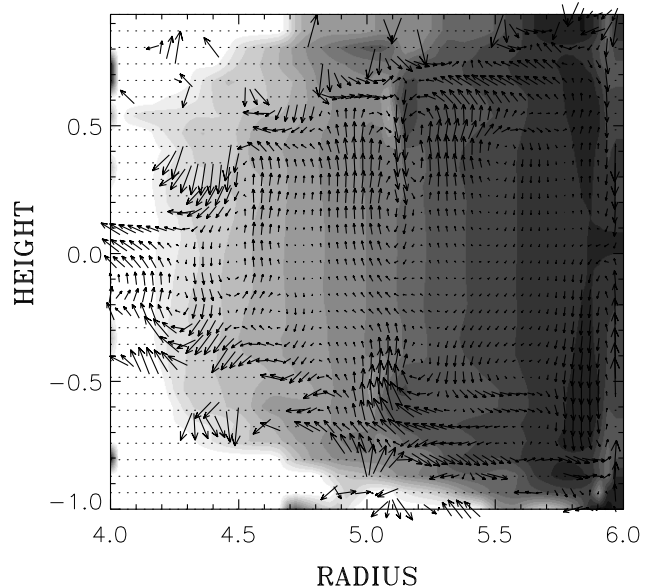, width=7.5cm}
  \end{center}
\caption{Vertical cut through the disk of Model II after 
15.7~orbital revolutions, excluding the high-velocity coronal
component of the disk. The shading represents the azimuthal component,
the arrows are the velocity vectors projected onto the $(r,z)$-plane
\label{cut_vel}}
\end{figure}

\begin{figure}
  \begin{center}
    \epsfig{file=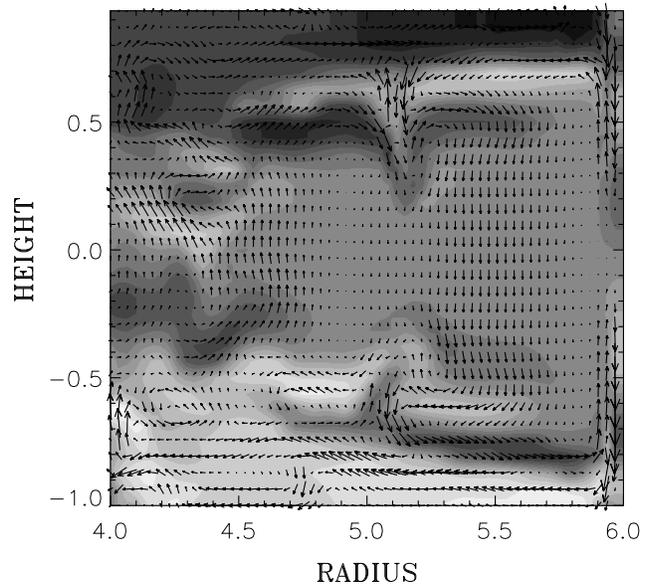, width=7.5cm}
  \end{center}
\caption{Vertical cut through the disk of Model II after 
15.7~orbital revolutions with an azimuthal-component shading and poloidal
magnetic-field vectors  \label{cut_mag}}
\end{figure}


\section{Patterns, transport, spectra}
\subsection{Glancing at the solutions}
Typical vertical slices from a run of Model~II are shown in
Figures~\ref{cut_vel} and \ref{cut_mag}. The central mass is located on the 
left side, outside the computational domain. The shadings
represent the azimuthal components of the magnetic and velocity
fields, while the arrows are the $(r,z)$-projections of the
actual vectors. Shaded areas near the equatorial plane of the disk
show the typical Keplerian velocity profile. However, at high
altitudes above the equator, a strong super-rotation emerges
coupled with relatively high infall velocities. (These high
radial velocities have been cut in Figure~\ref{cut_vel} to better
show the velocities in the disk.)

The computational
domain thus covers part of the corona of the disk where the
density falls below $10^{-3}$ times the equatorial density.
The infall velocity in the corona is of the same order of magnitude
as the Keplerian velocity, whereas the non-orbital velocities
within the disk are much smaller. The actual flow in the disk
as shown in Figure~\ref{cut_vel} is limited
by $0.35 c_{\rm ac}$. 

After saturation of the turbulence, the temporal behavior
of $\alpha_{\rm SS}$ changes. The fairly smooth function turns into 
an oscillatory behavior. The typical feature of this moment
is the zero-crossing of the Reynolds stress which is essentially
negative all through the saturation parts of the simulations.
It is expected that the infall of matter at the outer radial
boundary tends to perform relaxation motions, in particular
for Models II--IV where matter falls in homogeneously. The
exclusion of the outermost six cell planes from the computation
of the spatially average $\alpha_{\rm SS}$ does not alter
the result (the actual relaxation motions are observed only
in the outermost three cell planes). Relaxation motions will 
consist mainly of $u_z$ components which do not affect radial
angular-momentum transport directly.

A problem connected with the homogeneous feeding of the disk
from the outer boundary is the appearance of a `density front'.
This is the reason why the density at the outer boundary is 
highest which may look odd in the context of an accretion disk. 
This incoming density enhancement can be partly leveled
by the presumed infall velocity at the outer boundary.


\subsection{Angular momentum transport}
According to the standard model of a turbulent disk according
to Shakura \& Sunyaev (1973), the $r\phi$-element of the 
stress tensor of velocity and magnetic-field fluctuations,
\begin{equation}
  W_{r\phi} = \frac{{\left\langle{\rho u_r' u_\phi' -
    \frac{1}{4\pi}B_r' B_\phi'}\right\rangle}}{\langle\rho\rangle},
\end{equation}
scales like the speed of sound such as
\begin{equation}
  W_{r\phi}=\alpha_{\rm SS} c_{\rm ac}^2,
\end{equation}
parameterized by an unknown $\alpha_{\rm SS}$ which is smaller than unity
in the case of subsonic turbulence. We normalize the stress by the average
density in the computational domain; other authors used the density
averaged in the equatorial plane leading to lower limits for $\alpha_{\rm SS}$.
This stress measures the angular
momentum transport in the disk; positive values mean outward transport.
We average this quantity over the entire computational domain according
to
\begin{equation}
  \frac{\int \rho u_r' u_\phi' r d\phi dz dr}{\int r d\phi dz dr}
\end{equation}
for the velocity and 
\begin{equation}
  \frac{1}{4\pi}\frac{\int B_r' B_\phi' r d\phi dz dr}{\int r d\phi dz dr}
\end{equation}
for the magnetic field fluctuations. As the azimuthal velocity contains
a large supersonic axisymmetric mode -- the Keplerian flow -- we adopt
$u_\phi' = u_\phi - \sqrt{GM/r}$ for the azimuthal velocity fluctuations. The
averaging is in fact carried out on a discrete computational mesh by
\begin{equation}
  \frac{\sum_j r_j \sum_i \sum_k \rho u_r' u_\phi'}{n_i n_k \sum_j r_j}
\end{equation}
for the velocity part and
\begin{equation}
  \frac{1}{4\pi}\frac{\sum_j r_j \sum_i \sum_k B_r' B_\phi'}{n_i n_k \sum_j r_j}
\end{equation}
for the magnetic part.

\begin{figure}[ht]
  \begin{center}
    \epsfig{file=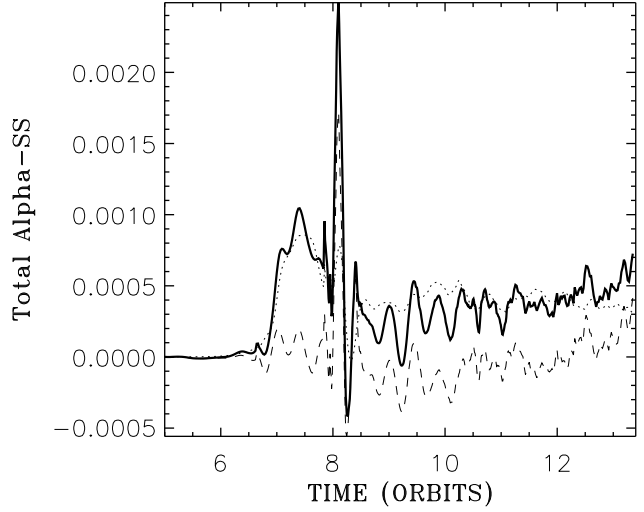, width=8.8cm}
  \end{center}
\caption{Angular momentum transport by Reynolds (dashed) and Maxwell 
(dotted) stresses as derived from a run of Model Ia, that is with
outflow boundary condition at the inner radial face and a closed
boundary for flow and magnetic fields at the outer radial face. 
The total transport (thick solid line) is positive 
and of the order of 0.0005.\label{plotalpha_temp2}}
\end{figure}

\begin{figure}[ht]
  \begin{center}
    \epsfig{file=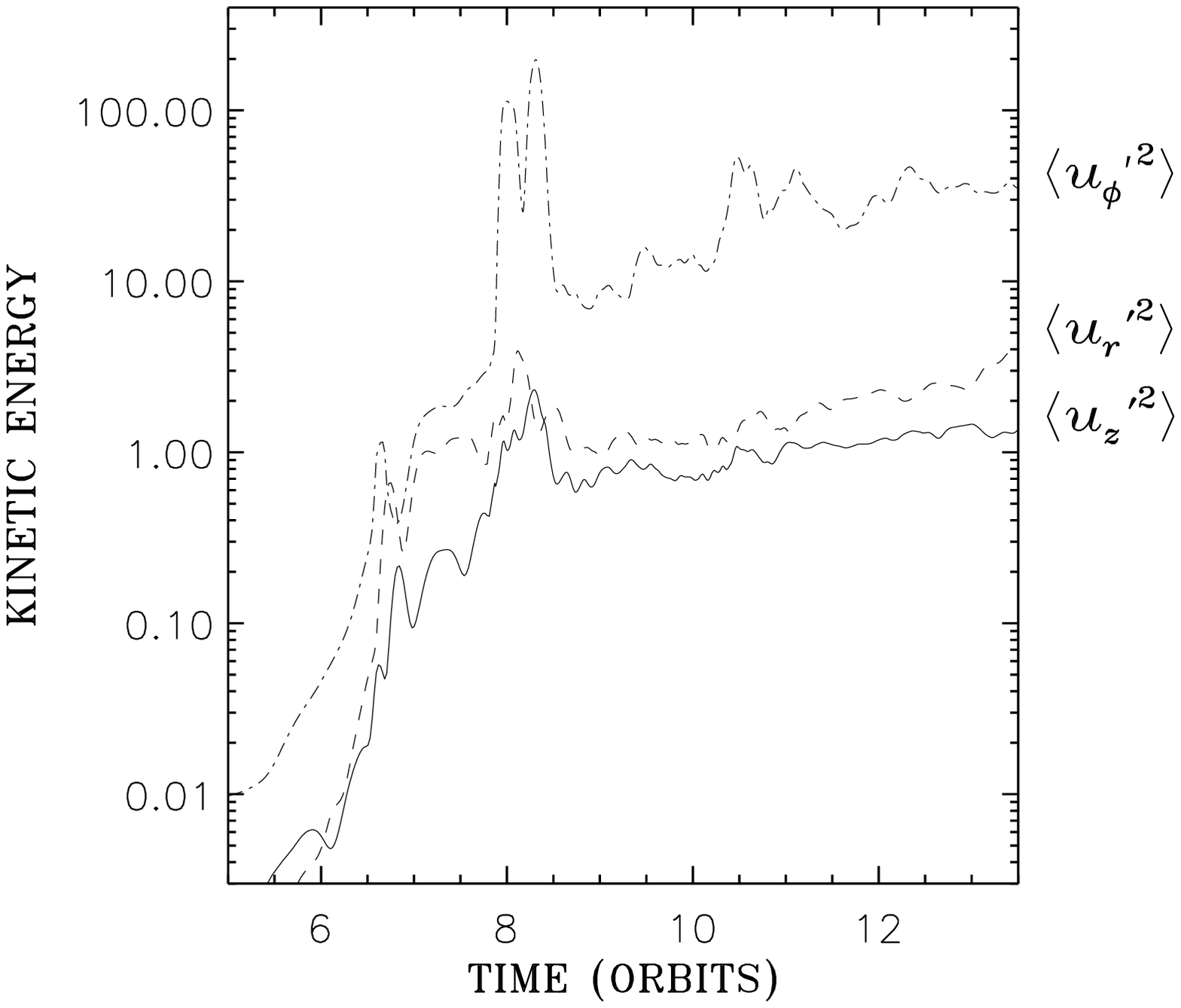, width=7.0cm}
  \end{center}
\caption{Energy in individual velocity components of Model Ia
\label{velo_temp2}}
\end{figure}

\begin{figure}[ht]
  \begin{center}
    \epsfig{file=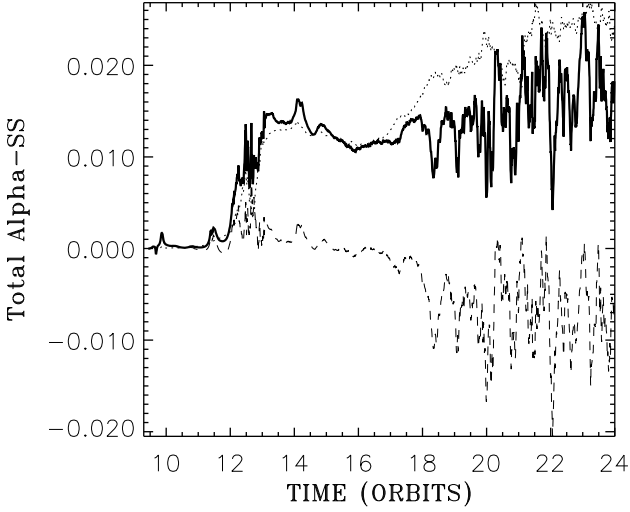, width=8.8cm}
  \end{center}
\caption{Angular momentum transport by Reynolds (dashed) and Maxwell 
(dotted) stresses as derived from a run of Model II. 
The total transport (thick solid line) is positive 
and of the order of 0.01\label{plotalpha_cs10bh}}
\end{figure}

\begin{figure}[ht]
  \begin{center}
    \epsfig{file=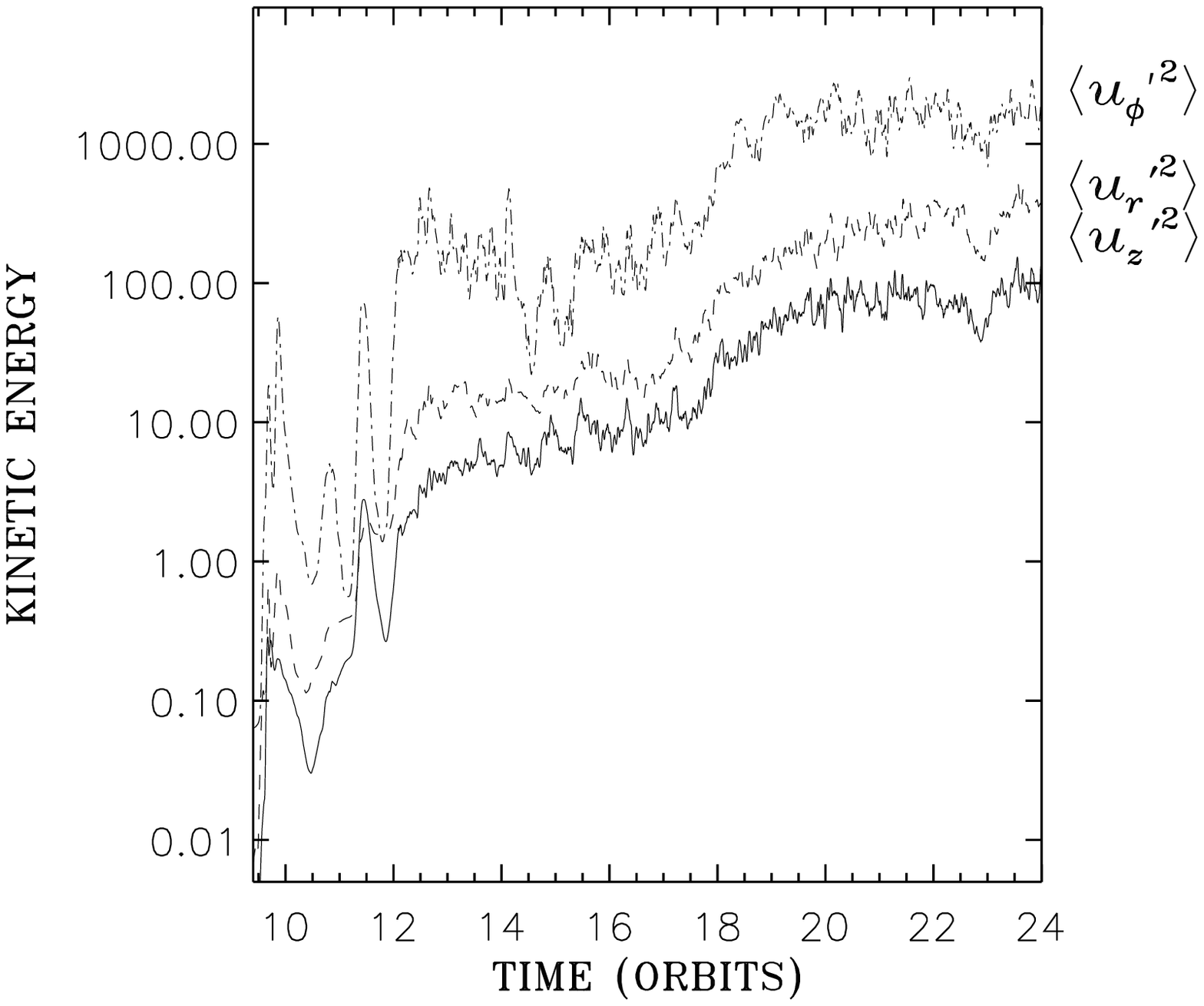, width=7.0cm}
  \end{center}
\caption{Energy in individual velocity components of Model II
\label{velo_cs10bh}}
\end{figure}

\begin{figure}[ht]
  \begin{center}
    \epsfig{file=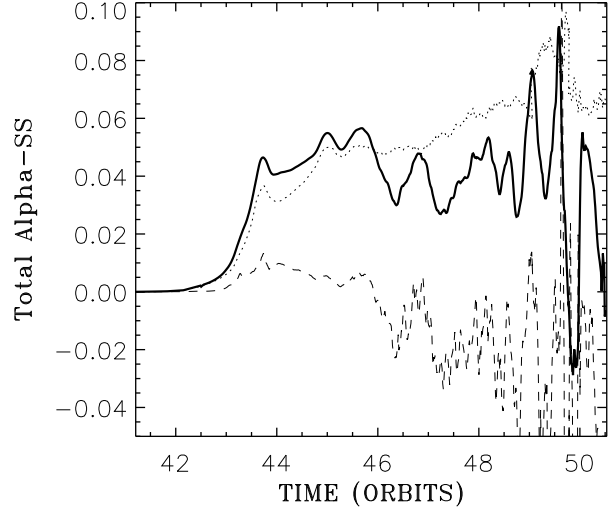, width=8.8cm}
  \end{center}
\caption{Angular momentum transport by Reynolds (dashed) and Maxwell 
(dotted) stresses as derived from a run of Model III. The homogeneous
inflow velocity is ten times higher in this simulation that in 
Figure~\ref{plotalpha_cs10bh}
\label{plotalpha_v01}}
\end{figure}

The time series of the outflow model Ia is shown
in Fig.~\ref{plotalpha_temp2}. Model Ia
in particular shows the typical short-lived occurrence of an
equivalent to the channel solution as was found in local box simulations;
a similar but less pronounced behavior is found in Model~Ib.
This period is the only time in both Models Ia and Ib when 
Reynolds stresses dominate the Maxwell stresses.

Time series for long runs of Models II and Model III are shown in
Figure~\ref{plotalpha_cs10bh} and \ref{plotalpha_v01}. 
Generally, the accretion model shows stronger angular momentum
transport than the the simpler inner-outflow Model Ia. 
Models~Ia and Ib run out of matter after about 8~orbital revolutions
with enabled magnetic fields, and the $\alpha_{\rm SS}$
values are no longer meaningful. 

A striking feature of the accretion Models II and III is the
change of sign in the Reynolds stress after about 4--6 orbits.
We find that, in a long run, the outward angular momentum transport
is a sole consequence of the magnetic stresses. Additionally,
the Reynolds part shows strong amplitude variations unlike the
Maxwell stress.

Figures~\ref{velo_temp2} and \ref{velo_cs10bh} show the respective temporal
evolution of the fluctuative kinetic energies. For the $z$- and
$r$-components, the zero mode is subtracted such as
$\langle u_z'^2\rangle=\langle u_z^2\rangle-\langle u_z\rangle^2$;
the azimuthal fluctuative energy is found by
$\langle u_\phi'^2\rangle=\langle (u_\phi - \sqrt{GM/r})^2\rangle$.
All three models show a clear energy `sorting' of the $z$-, $r$-, and
$\phi$-components in this order.

A particularly long run was performed with a computational domain
only half as large as Models II and III and is denoted by IV. The
initial and boundary conditions are like in Model II. The 
long-term behavior is not too similar to that of Model II with
very strong oscillations of Reynolds versus Maxwell stresses.
A behavior comparable with Model~II may be spotted during the
first 8~orbital periods. After that time, the energy sorting
of the radial and vertical components is given up. 
All models from Ia to IV appear to be not fully saturated with
respect to the ever-growing kinetic energies.

\begin{figure}[ht]
  \begin{center}
    \epsfig{file=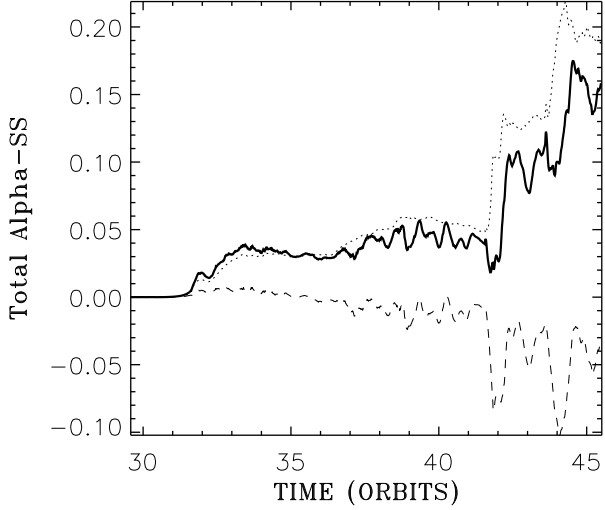, width=8.8cm}
  \end{center}
\caption{Angular momentum transport by Reynolds (dashed) and Maxwell 
(dotted) stresses as derived from a run of Model V. Density is
not falling in homogeneously, but with a predefined Gaussian vertical
distribution; the (homogeneous) inflow velocity is 0.01
\label{plotalpha_bgauss}}
\end{figure}

\begin{figure}[ht]
  \begin{center}
    \epsfig{file=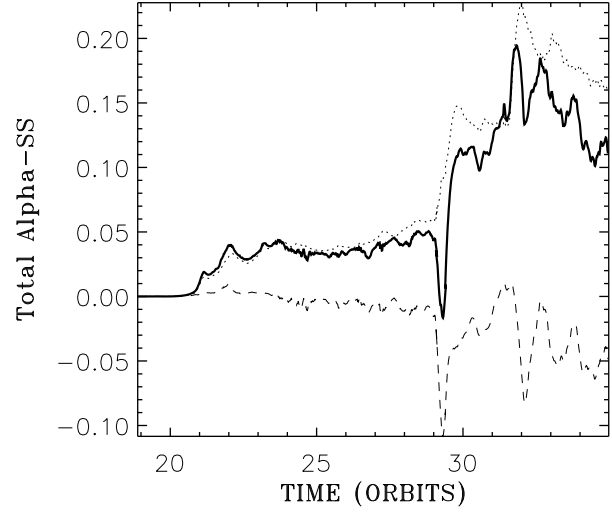, width=8.8cm}
  \end{center}
\caption{Angular momentum transport by Reynolds (dashed) and Maxwell 
(dotted) stresses as derived from a run of Model VI. The inflow
velocity is ten times higher than in model V, 
Figure~\ref{plotalpha_bgauss}
\label{plotalpha_bgv01}}
\end{figure}

We should note that a homogeneous infall might be suitable for
a realistic disk in the sense that the infalling matter does not
`know' the structure of the disk. Nevertheless, numerical
problems may occur when the incoming homogeneous 'matter front'
meets the near-Gaussian density structure. Models V and VI
apply an inflow with pre-defined Gaussian infall profile over
$z$. The respective $\alpha_{\rm SS}$-profiles are shown in
Figures~\ref{plotalpha_bgauss} and \ref{plotalpha_bgv01}, the
first having an infall velocity of $u_{\rm in}=-0.01$, the second
$u_{\rm in}=-0.1$. The initial behavior is similar to the respective
Models II and III including the change of sign of the Reynolds stress
after a couple of orbits. After as many as 10~orbital periods,
the disk turns into a significantly different regime with 
enhanced outward transport in both models.

\begin{figure}[ht]
  \begin{center}
    \epsfig{file=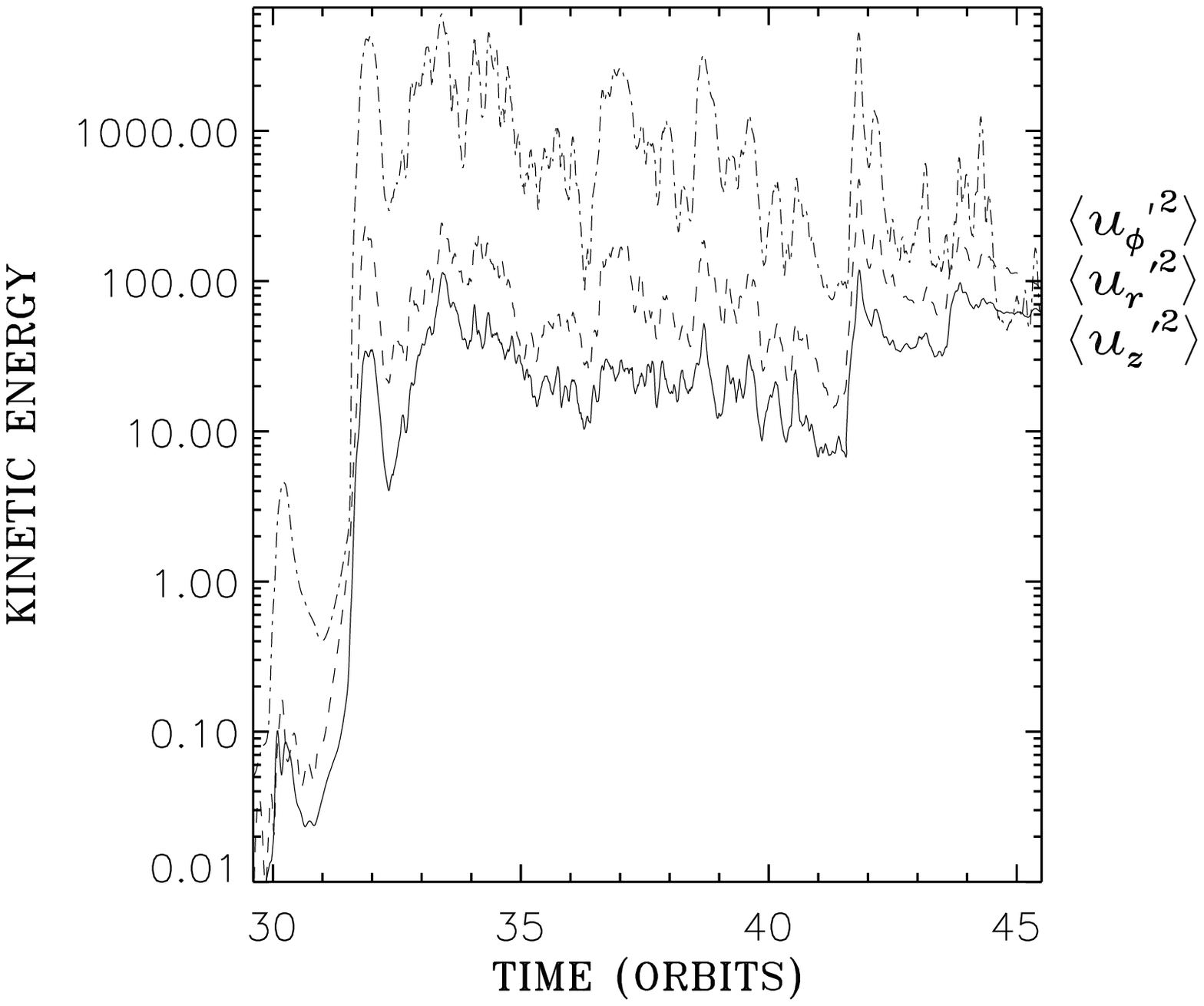, width=7.0cm}
  \end{center}
\caption{Energy in individual velocity components of Model V
\label{velo_bgauss}}
\end{figure}

\begin{figure}[ht]
  \begin{center}
    \epsfig{file=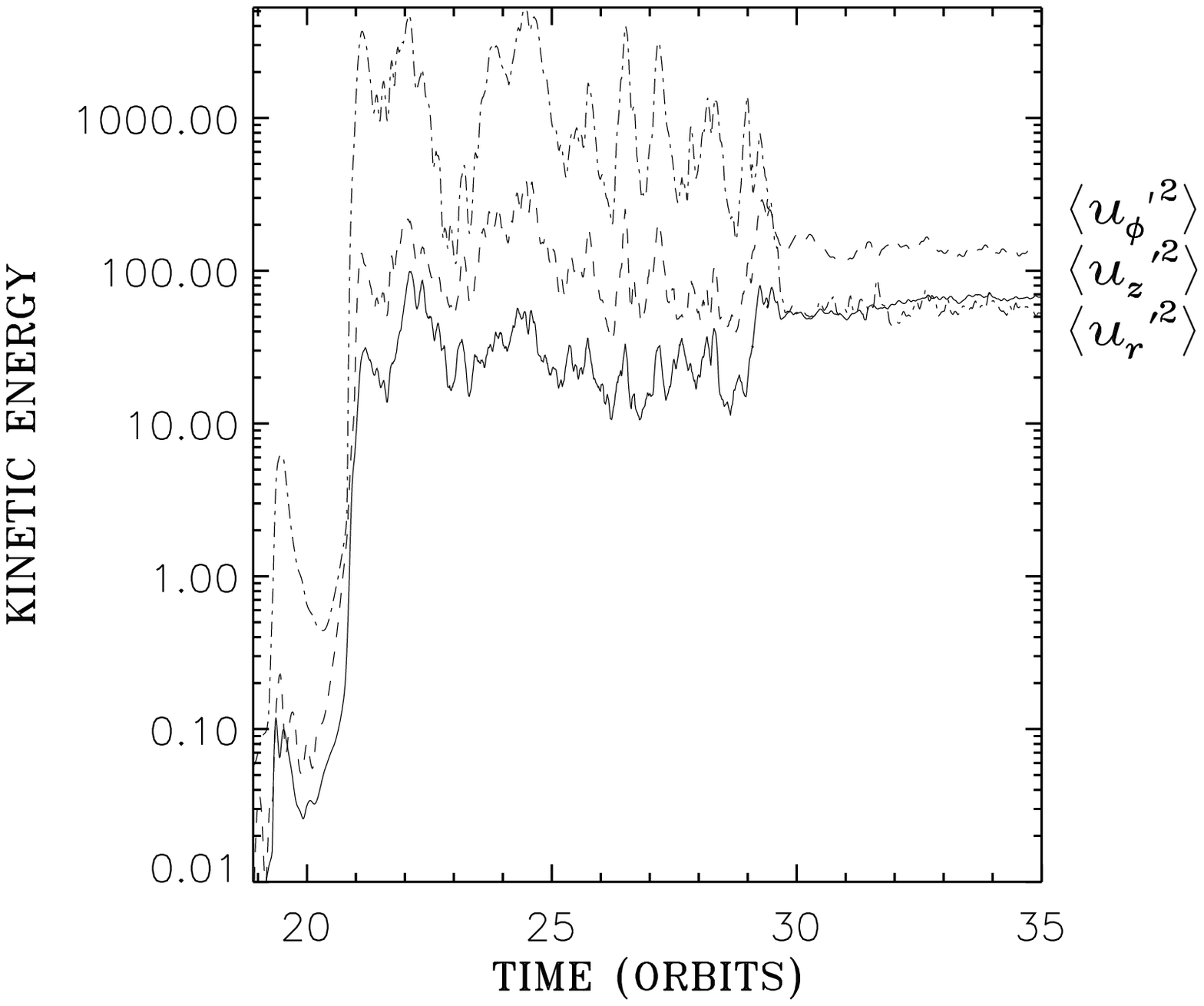, width=7.0cm}
  \end{center}
\caption{Energy in individual velocity components of Model VI
\label{velo_bgv01}}
\end{figure}

The respective kinetic energies are given in Figures~\ref{velo_bgauss}
and \ref{velo_bgv01}. In contrast to the above studied models,
energies do not develop in an ever-growing way, but appear to fluctuate
about an average. The only exception is the regime change near
$t=42\,T_{\rm orb}$ (Model V) and $t=29\,T_{\rm orb}$ (Model VI)
when the average (essentially of $\langle u_z'^2\rangle$) changes, but
the stationary appearance remains.

\begin{table}
\caption{\label{tab2}Results of the models given in Table~\ref{tab1}.
The duration is the total period of the run with enabled magnetic fields.
The last column gives the period from which the $\alpha_{\rm SS}$-average 
was obtained. Times are in units orbital revolutions of the inner edge}
\begin{small}
\begin{tabular}{lllcc}
\hline
Mod.&Resolution&Duration& $\langle\alpha_{\rm SS}\rangle$ & during \\
\hline
Ia   & $31\times61\times351$ & \phantom{0}8.3 & 0.0003 & \phantom{0}8.5--13.4 \\
Ib   & $31\times61\times351$ & 12.2 & 0.01--0.02 & 8.0-13.5\\
II   & $31\times61\times351$ & 14.7 & 0.014\phantom{0} & 13.0--24.0\\
III  & $31\times61\times351$ & \phantom{0}9.7 &0.041\phantom{0}& 44.0--50.9 \\
IV   & $31\times31\times351$  & 41.5 & 0.014\phantom{0}& 35.0--69.8\\
V    & $31\times61\times351$  & 11.9 & 0.038\phantom{0}& 33.0--41.5\\
''   &    ''                  &      & 0.12\phantom{00}& 42.0--45.5 \\
VI   & $31\times61\times351$  & 16.1 & 0.037\phantom{0}& 22.0--29.0 \\
''   &    ''                  &      & 0.14\phantom{00}& 30.0--35.0 \\
VII  & $31\times31\times151$  & 36.8 & 0.080\phantom{0}& 32.0--44.0 \\
VIII & $31\times61\times351$  & 22.4 & 0.07\phantom{00}& 29.0--47.1 \\
\hline
\end{tabular}
\end{small}
\end{table}

\begin{figure}[ht]
  \begin{center}
    \epsfig{file=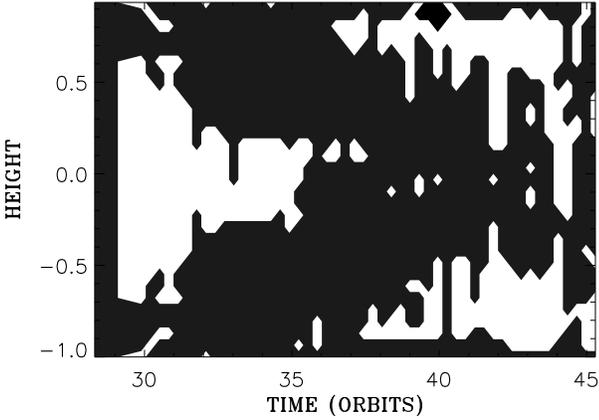, width=8.8cm}
  \end{center}
\caption{\label{alptime_bgauss}Distribution of the total stress
$\alpha_{\rm SS}$ versus time and vertical coordinate for Model V. 
Light areas represent inward transport, dark areas outward transport}
\end{figure}

The distribution of $\alpha_{\rm SS}$ versus time and the $z$-direction
is shown in Figure~\ref{alptime_bgauss} for Models V. Outward angular momentum 
transport is marked with white areas, inward transport by dark areas.
Strong transport is observed at large distances from the equator
shortly after the onset of the instability. Regions of outward
transport migrate towards the equator. This is particularly marked
in the Model-V plot for which we found a development into a 
strong-transport regime (cf.\ Figure~\ref{plotalpha_bgauss}).
Positive angular momentum transport is not restricted to high 
altitudes anymore.

\begin{figure}[ht]
  \begin{center}
    \epsfig{file=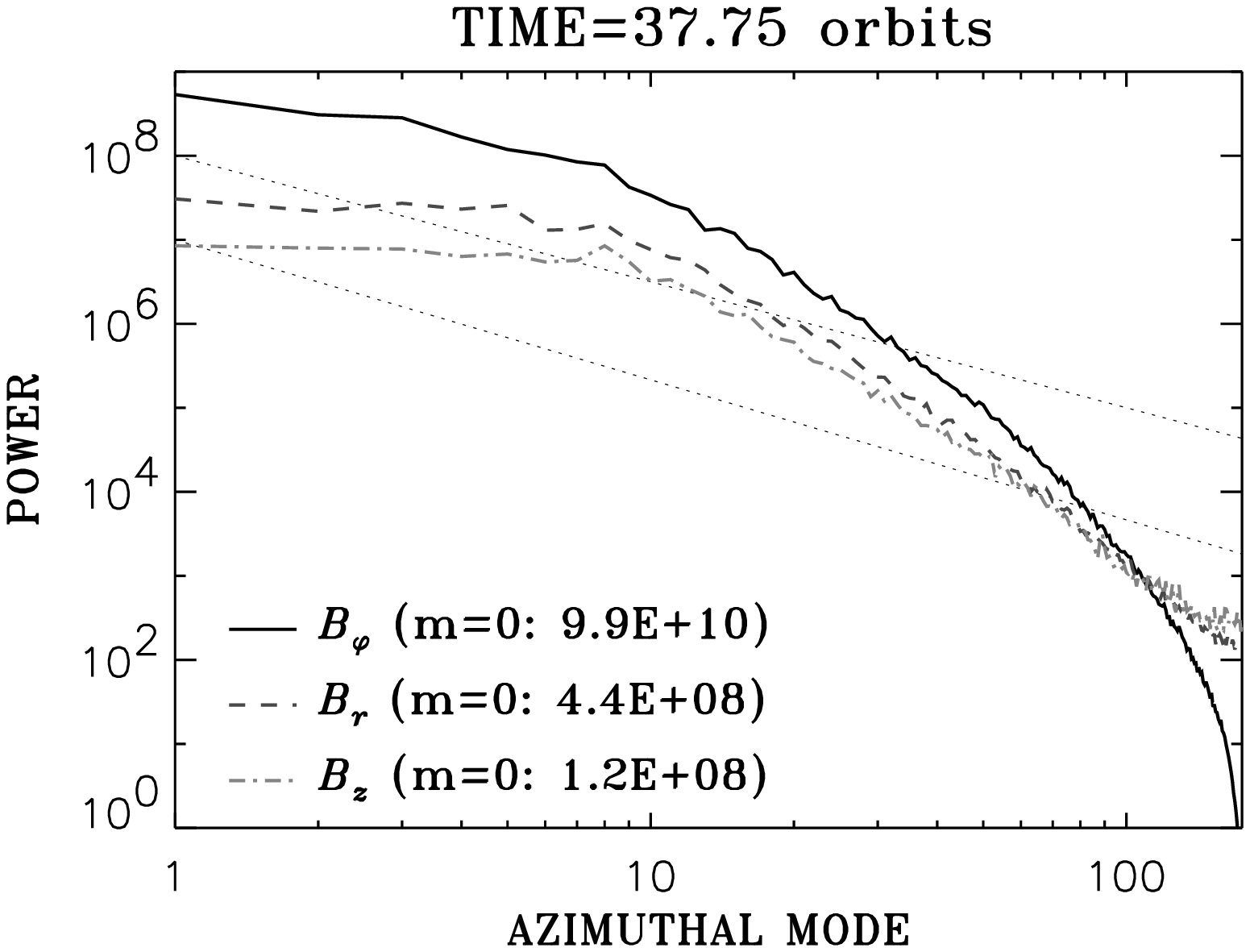, width=8.8cm}
  \end{center}
\caption{Azimuthal Fourier decomposition of the velocity components
after 37.75 orbital periods in Model V\label{fourcolor_bgauss35}}
\end{figure}

\begin{figure}[ht]
  \begin{center}
    \epsfig{file=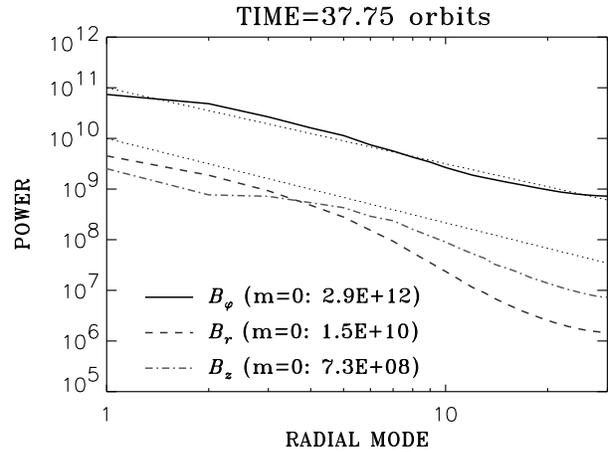, width=8.8cm}
  \end{center}
\caption{Radial Fourier decomposition of the magnetic-field
         components after 37.75 orbital periods in Model V
\label{fourcolor_bgauss35r}}
\end{figure}

\begin{figure}[ht]
  \begin{center}
    \epsfig{file=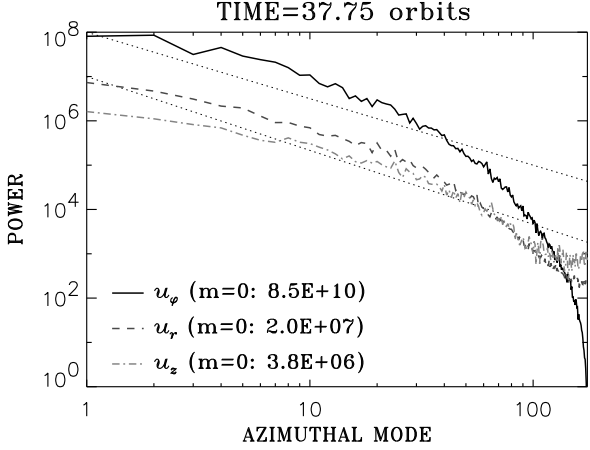, width=8.8cm}
  \end{center}
\caption{Azimuthal Fourier decomposition of the velocity
         components after 37.75 orbital periods in Model V
\label{fourcolor_bgauss35v}}
\end{figure}

\subsection{Spectral decomposition}
Power spectra resulting from Fourier transforms of the magnetic field
of Model V over the the azimuthal and radial directions are shown 
in Figures~\ref{fourcolor_bgauss35} and \ref{fourcolor_bgauss35r},
respectively. The azimuthal decomposition of the velocity field
is shown in Fig.~\ref{fourcolor_bgauss35v}. The individual
azimuthal spectra for all points in the ($z,r$)-plane were 
averaged, as were the points of the ($\phi,z$)-plane for the 
radial spectra. In the azimuthal spectra, we observe
slight maxima in power at $k=5$ and $k=8$, followed by a strong, roughly
power-law decline between $k\sim 20$ and $k\sim 100$. The magnetic
spectrum is steeper than the kinetic one. The underlying power-laws
would be $E\propto k^{-3}$ to $E\propto k^{-4}$ which is significantly
steeper than a Kolmogorov spectrum, and is similar to what Armitage
(1998) found for a global simulation in the far-$k$ range. The Kolmogorov
spectrum of isotropic turbulence would exhibit a wavenumber 
exponent of $-5/3$; here we face MHD turbulence which implies an 
effect of Alfv\'en waves impeding the transfer of energy towards 
smaller scales. The energy spectrum of isotropic MHD turbulence
is $E\propto k^{-3/2}$, whence more shallow than the Kolmogorov spectrum.
Contrasting with the azimuthal decomposition, the radial spectrum
matches the theoretical MHD spectrum over the entire range
of modes as can be seen in Fig.~\ref{fourcolor_bgauss35r}.

\begin{figure}
    \epsfig{file=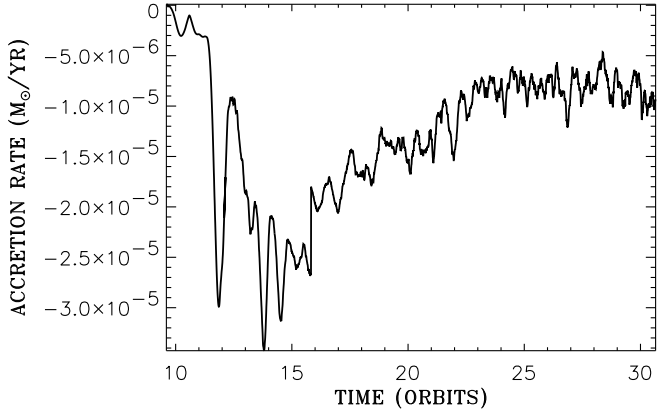, width=8.6cm}
\caption{Accretion rate of Model~II scaled to a system at 
100~AU distance from the central object.
\label{plotoutflux_cs10bh}}
\end{figure}

\subsection{Accretion rate}
Choosing the rough dimensions of an inner protostellar system
of 100~AU radius for scaling the mass flux through our models,
we obtain an accretion rate in physical units as shown in
Fig.~\ref{plotoutflux_cs10bh}. Widely accepted values for the
accretion rates of protostellar disks around solar-type
stars are $10^{-7}$ to $10^{-6}\,M_\odot$/yr. The compilation
of results for those stars by Strom (1994) rather favors
$10^{-6}\,M_\odot$/yr. The accretion rate increases with
stella mass and may already exceed $10^{-5}\,M_\odot$/yr
for a 3-solar-mass star. Our results such as given in
Fig.~\ref{plotoutflux_cs10bh} are thus not too
far from the accretion rates deduced from the observations.


\subsection{Comparison with other studies}



Among stratified local simulations, we find a number of very similar results
as regards the magnitude of $\alpha_{\rm SS}$, such as in
Stone et al. (1996) who obtained $0.01$,
Hawley et al. (1996) where an average of 0.016 was found, with typical 
values of 0.01. The largest box used delivered 0.02, the smallest, 0.003.
Upon studying the dependence of the turbulence on the rotation law
$\Omega\propto r^{-q}$, Hawley et al. (1999) found $10^{-2}$ for $q=1.5$.
A most recent local-box simulation by Ziegler \& R\"udiger (2000b)
covering 288~orbital periods
resulted in a space-time average of 0.015, where the Reynolds part
was $2.8\cdot 10^{-3}$ and the Maxwell part was $1.2\cdot 10^{-2}$.
The stress was normalized with the averaged equatorial-plane
pressure and thus provides a lower limit for $\alpha_{\rm SS}$.
The early stratified-box simulations by Brandenburg et al. (1995)
provided $\alpha_{\rm SS}$ an order of magnitude lower between 0.001
and 0.005. All the models agree upon the dominance of Maxwell
stresses over Reynolds stresses.

It is most interesting to note that other global simulation such as
performed by Armitage (1998) or Hawley (2000) result in an order
of magnitude higher values of 0.2 to 0.3. Contrasting with this,
the global spherical, but unstratified simulation by Drecker (2000) 
gives again an $\alpha_{\rm SS}$ which is an order of magnitude
lower than shown here, i.e.\ $10^{-3}$. Those simulations alone
included a physical viscosity. The simulations of Hawley (2000)
-- and with somewhat different objectives, those of Machida et al.\ (2000)
-- had open boundaries in radial and vertical directions. This
fact limits the run-time of the models which are prone to running
out of matter.

\section{Dynamo effect}
\begin{figure}[ht]
  \begin{center}
    \epsfig{file=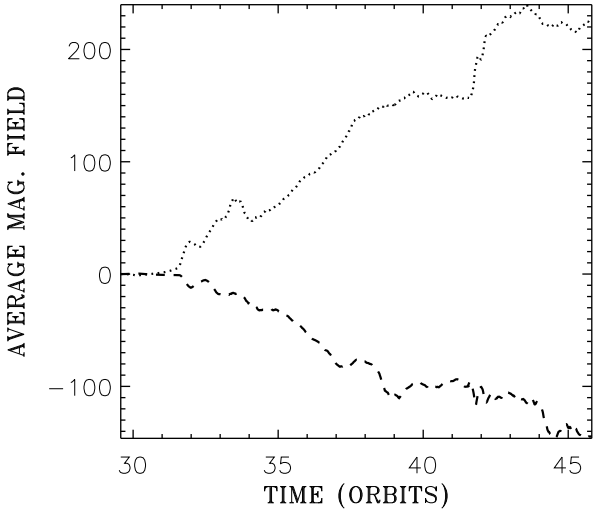, width=8.8cm}
  \end{center}
\caption{\label{bphi_bgauss}Average toroidal magnetic field of Model~V.
The dashed line is the toroidal field averaged only in the northern 
hemisphere, the dotted line is from the southern hemisphere}
\end{figure}

The long-term existence of non-zero toroidal fields in each hemisphere
as shown for example in Figure~\ref{bphi_bgauss} for
Model V indicates the possible action of a dynamo in the disk.

The local-box simulations by Hawley et al.\ (1996) impose severe
constraints on the growth of an average magnetic field because 
of the periodicity of the boundary conditions. A non-zero term 
of the average magnetic-field growth is due to the shear and requires
the presence of an average radial component, in the local nomenclature
$\langle B_x\rangle$. We follow their derivation of the averages
by replacing the volume curl integral by a vector-produc surface
integral of (3). The average
\begin{equation}
  \frac{\partial\langle\vec B\rangle}{\partial t} =
  \frac{1}{\pi\Delta r^2\Delta z}
  \int\limits_V \rot (\vec u\times \vec B - \eta \rot \vec B) dV
\end{equation}
is equivalent to
\begin{equation}
  \frac{\partial\langle\vec B\rangle}{\partial t} =
  \frac{1}{\pi\Delta r^2\Delta z}
  \int\limits_S d\vec S\times (\vec u\times \vec B - \eta \rot \vec B)
\end{equation}
and delivers average components as
\begin{eqnarray}
  \frac{\partial\langle B_z\rangle}{\partial t} &=&
  -\left\langle{\frac{\partial}{\partial r}u_r B_z}\right\rangle\\
  \frac{\partial\langle B_r\rangle}{\partial t} &=&
\phantom{-}\left\langle{\frac{\partial}{\partial z} u_r B_z}\right\rangle
  +\eta\left\langle{\frac{\partial}{\partial z}\frac{\partial B_z}{\partial r}
  }\right\rangle\\
  \frac{\partial\langle B_\phi\rangle}{\partial t}&=&
  \phantom{-}\left\langle{\frac{\partial}{\partial z} u_\phi B_z}\right\rangle
  -\eta\left\langle{\frac{1}{r}\frac{\partial}{\partial z}
  \frac{\partial B_z}{\partial \phi}-\frac{\partial^2B_\phi}{\partial z^2}
  }\right\rangle-\nonumber\\
&&-\left\langle{\frac{\partial}{\partial r}u_\phi B_r}\right\rangle
  +\eta \left\langle{\frac{\partial}{\partial r}\frac{1}{r}
  \frac{\partial rB_\phi}{\partial r}}\right\rangle
\end{eqnarray}
These considerations show that slight initial fluctuations in
$\vec B$ are able to create non-zero average magnetic fields
at any time, even if their $\langle \vec{B'}\rangle$ vanishes,
as a consequence of shear and radial accretion through $u_{\rm in}$.
In particular, the non-constant vertical profile of the orbital
velocity produces a $B_\phi$ from the initial perturbation $B_z$.
Fluctuations in the accretion flow generate $B_r$ and $B_z$.
The actual radial differential rotation then quickly produces
$B_\phi$ from even very small $B_r$, and toroidal fields are prone
to dominate the disk fields.

\subsection{Sketching mean-field dynamos}
Kinematic dynamos have been extensively examined in numerous
previous publications concerning various types of geometries.
A widely used mean-field approach to circumnavigate the difficulties
with resolving the small scales in a model applies the likely
amplification of a magnetic field through a fluctuative electromotive
force parallel to the actual large-scale magnetic field. This
effect is generally expressed by the term `$\alpha$-effect' and
can be thoroughly studied in e.g.\ Krause \& R\"adler (1981).
In a model splitting large and small scales, the induction 
equation extends beyond the large-scale electromotive force
such as
\begin{equation}
{\partial \langle\vec{B}\rangle \over \partial t} = \rot
\left(\langle\vec{u}\rangle
\times \langle\vec{B}\rangle + \vec{\cal{E}}\right).
\end{equation}
The $\alpha$-effect is part of the development of the small-scale
electromotive force as
\begin{equation}
{\cal E}_i = \alpha_{ij}^{\rm dyn} \ \langle B_j\rangle - \eta_{ijk}
\langle B_k\rangle_{,j} +\dots,
\label{emf}
\end{equation}
where $\alpha_{ij}^{\rm dyn}$ and $\eta_{ijk}$ are tensors in general. If
written in components, the induction equation shows that the essential
component of the $\alpha^{\rm dyn}$ tensor is the $\alpha_{\phi\phi}^{\rm dyn}$
component
which converts azimuthal fields into radial and vertical fields,
whereas the differential rotation converts radial and vertical
fields back into the toroidal component by the large-scale part
of $\langle\vec{u}\rangle\times \langle\vec{B}\rangle$. This 
approximation is usually referred to as an $\alpha\Omega$-dynamo.

Second-order correlation approximation leads to a relation
of $\alpha^{\rm dyn}$ with the helicity of the flow,
\begin{equation}
 \alpha^{\rm dyn} = 
    -\frac{1}{3}\int\limits_0^\infty \Bigl\langle{\vec u'(t)\cdot
     \rot \vec u'(t-\tau)}\Bigr\rangle\,d\tau,
\end{equation}
which is, when approximated by a typical time-scale $\tau_{\rm corr}$,
\begin{equation}
\alpha^{\rm dyn}\sim -\langle \vec u'\cdot \rot \vec u'
      \rangle \,\tau_{\rm corr}.
\end{equation}
The same principle is applicable to the current helicity
(Keinigs 1983) giving
\begin{equation}
\alpha^{\rm dyn}\sim -\frac{\eta}{B^2}\langle \vec B'\cdot \rot \vec B'
      \rangle.
\end{equation}
Among other issues, the evolution of these quantities for 
the helicity will be subject of the following sections.

\subsection{Symmetry of magnetic fields}
Special attention is payed to the symmetry of the magnetic fields
of the solutions. In general, the preference of a certain symmetry
is a tool to check the consistency with mean-field dynamo models.
Additionally, dipolar (antisymmetric) fields are supposed to
support the launch of winds forming jets from the disk better
than quadrupolar (symmetric) fields, with the latter rather providing
closed field lines within the disk. The mean-field disk dynamo
of Rekowski et al.\ (2000) produced dipolar magnetic fields
for negative $\alpha^{\rm dyn}$ in the northern hemisphere.

A number of mean-field dynamo models in one to three dimensions
delivered varying symmetry of the solutions with a major influence
of the boundary conditions -- vacuum versus perfectly conducting.
Three-dimensional investigations of stability by Meinel et al.\ (1990)
and Elstner et al.\ (1992) found quadrupolar solutions for the
condition of low-conductivity surroundings of the disk, whereas
perfectly conducting boundaries delivered dipolar solutions.
A local model with periodic boundaries will thus not be able
to answer the question of the final symmetry of the solutions.
An ideal global model would comprise the entire investigated object
and is less depending on the surroundings (which will be vacuum
in very good approximation).

\begin{figure}
  \begin{center}
    \epsfig{file=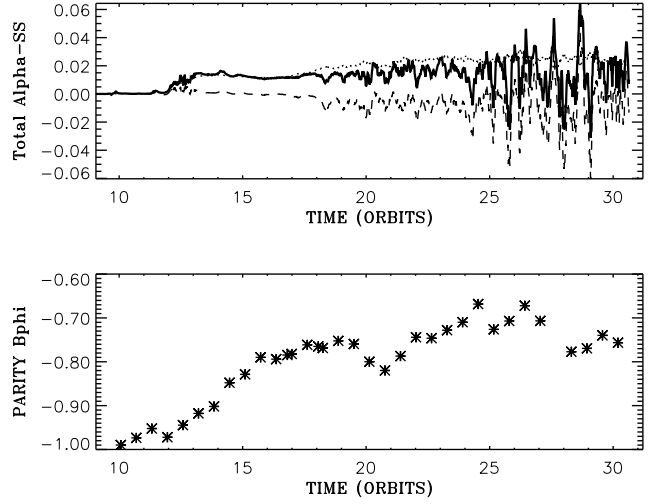, width=9.0cm}
  \end{center}
\caption{\label{parityphd_cs10bh}Total $\alpha_{\rm SS}$ and parity
for Model~II. A parity of $-1$ denotes fully antisymmetric
(dipolar) solutions}
\end{figure}

\begin{figure}
  \begin{center}
    \epsfig{file=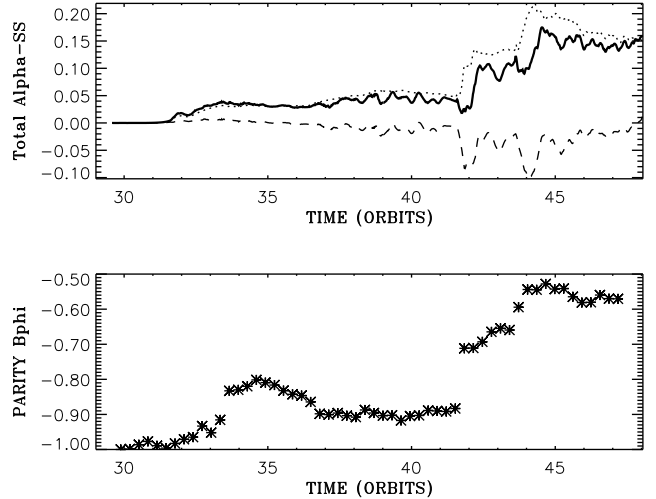, width=9.0cm}
  \end{center}
\caption{\label{parityphd_bgauss}Total $\alpha_{\rm SS}$ and parity
for Model~V.}
\end{figure}

The parity of a the toroidal magnetic field is measured by
\begin{equation}
  P=\frac{E_{\rm s}-E_{\rm a}}{E},
  \label{parity}
\end{equation}
where $E_{\rm s}$ and $E_{\rm a}$are the symmetric and antisymmetric
energy parts resp., and $E$ is the total energy of in toroidal
field component. 
Figures~\ref{parityphd_cs10bh} and \ref{parityphd_bgauss} show
the temporal development of parity of the toroidal magnetic
fields in the Models~II and V along with a repeated plot
of $\alpha_{\rm SS}$.

We should emphasize that the initial perturbation of the magnetic
field had dipolar symmetry. Since the total flux through the
vertical surfaces, however, vanishes, field configurations
with either symmetry may arise from the evolution of the model.
The dipolar initial perturbation may not have been fully
re-organized as the diffusive time-scale
\begin{equation}
  \tau_{\rm diff} = H^2/\eta
  \label{tau_diff}
\end{equation}
is 1600 rotation periods for the low-diffusivity model~II and
160 revolutions for the high-diffusivity model~V. The latter 
covered $0.12\,\tau_{\rm diff}$. An additional experiment 
used a completely mixed initial magnetic-field perturbation
according to a distribution
which has zero parity in $B_z$, and the toroidal component immediately
emerging from the slight vertical has zero parity, too. Again, the
magnetic flux through the vertical surfaces is zero. Such initial 
conditions have been applied to Models~II and V. The runs lasted
roughly 10~revolution resulting in parity variations with a slight
tendency towards quadrupolar (symmetric) solution ($P$ from +0.1
to +0.2).

\subsection{The dynamo of the disk model}
Average kinetic helicities were computed based on a layer
slightly below one density scale-height in both hemispheres. 
The average is taken in $\phi$- and $r$-direction. There is
a considerable scatter in the time series of the helicity,
but temporal averages are all negative on the northern
side for all models. The only exception is the northern
part of Model~II with a near-zero value. Negative sign
also holds for the two test models with mixed-parity initial
perturbation. Average current helicities are also
negative on the northern side (and positive on the southern one)
throughout almost all
models. The only exceptions are the mixed-parity models
which show a current helicity with opposite sign compared
with the kinetic helicity in both hemispheres.

\begin{figure}[ht]
  \begin{center}
    \epsfig{file=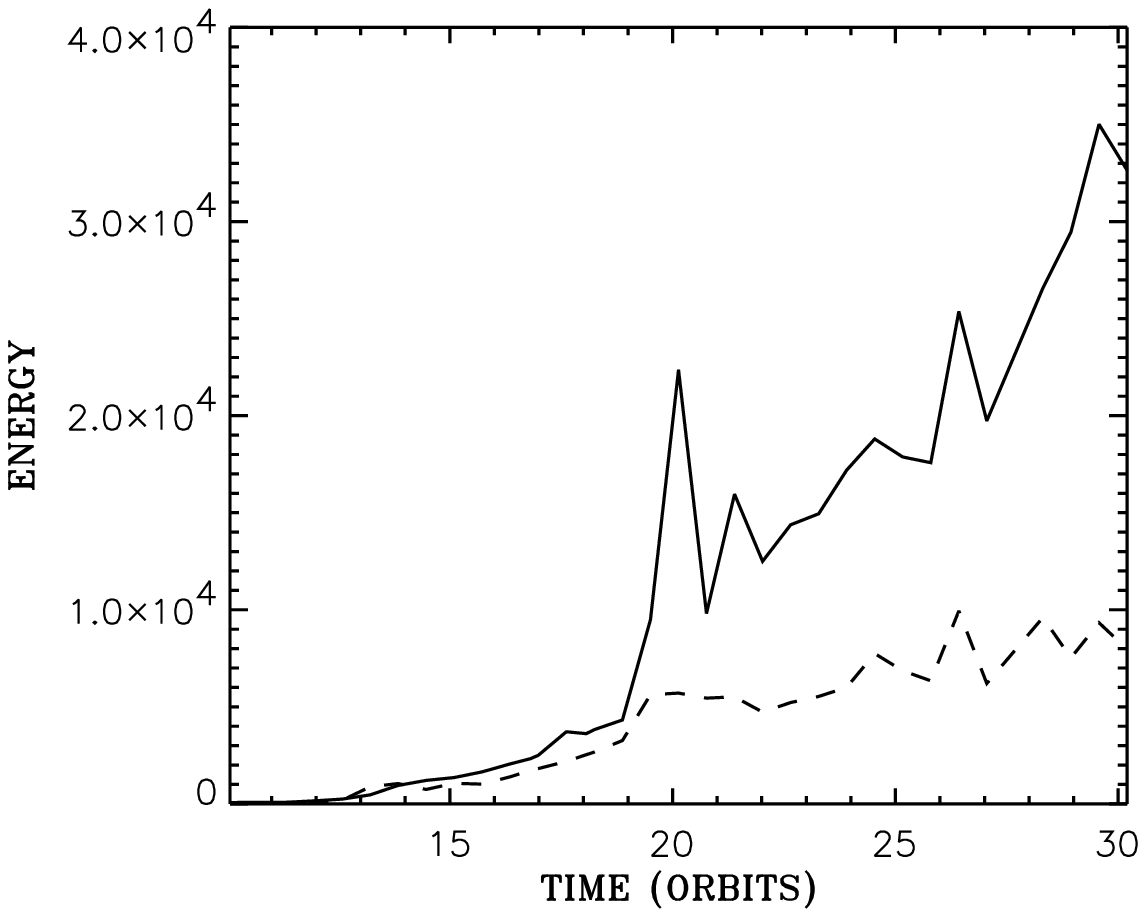, width=8.8cm}
  \end{center}
\caption{\label{helres_paper_cs10bh}
Temporal evolution of kinetic (solid line) and
magnetic (dashed line) energies of Model~II.
The energy is measured as a radial and azimuthal total in a 
horizontal layer at $z=+0.39$ which
is slightly below one density scale-height}
\end{figure}

\begin{figure}[ht]
  \begin{center}
    \epsfig{file=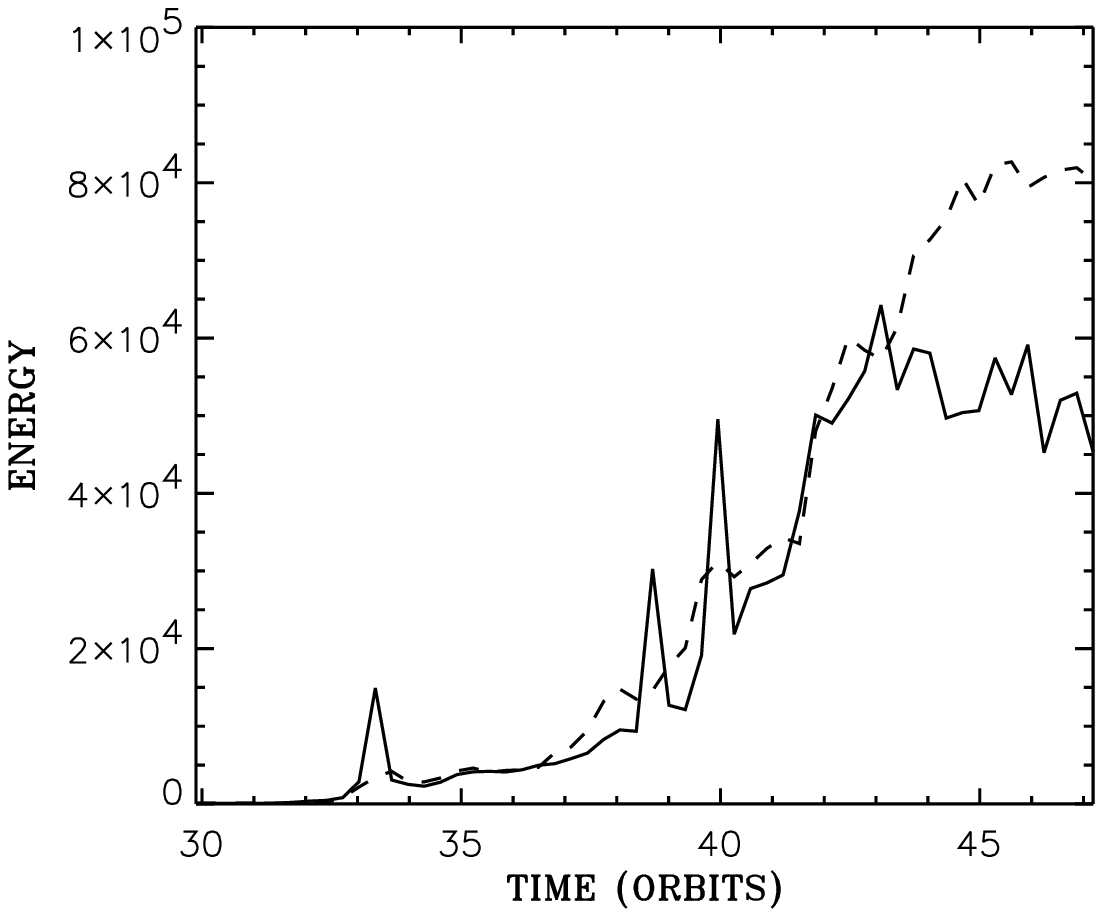, width=8.8cm}
  \end{center}
\caption{\label{helres_paper_bgauss}
Temporal evolution of kinetic (solid line) and
magnetic (dashed line) energies of Model~V.
The energy is measured as a radial and azimuthal total in a 
horizontal layer at $z=+0.39$ which
is slightly below one density scale-height}
\end{figure}

Since the angular momentum transport is dominated by
magnetic stresses, we also ask about the {\it energy\/} in
the flow fluctuations and the magnetic field. The temporal evolutions
of these two quantities in Figures~\ref{helres_paper_cs10bh} and
\ref{helres_paper_bgauss} show a clear difference between
the Models~II and V. Kinetic-energy dominance holds for
Model~II, whereas magnetic dominance is found for Model~V.

\begin{figure}[ht]
  \begin{center}
    \epsfig{file=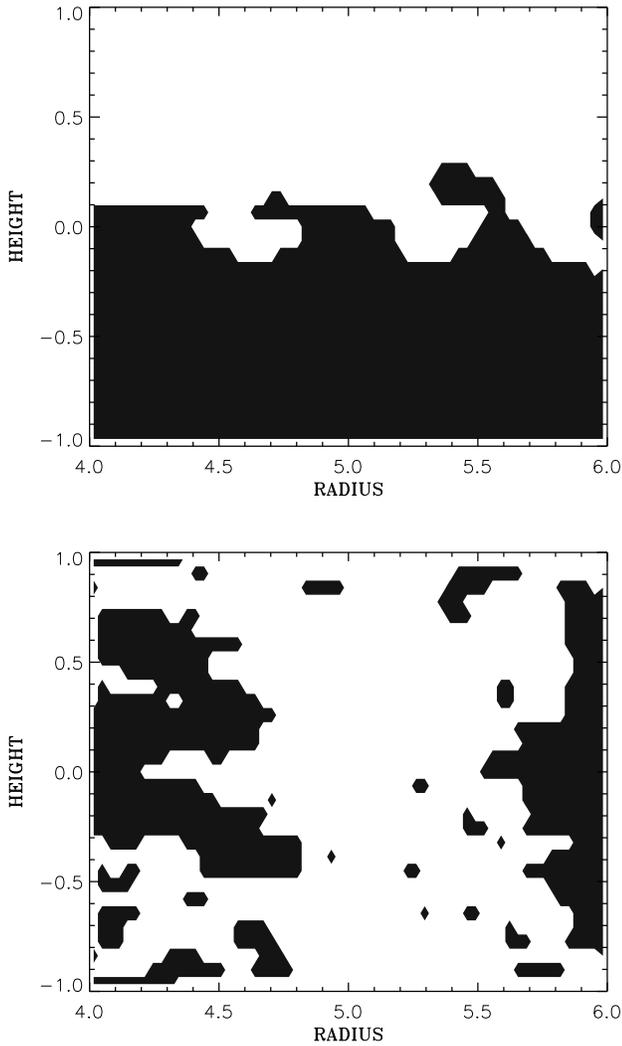, width=8.8cm}
  \end{center}
\caption{\label{emf_bgauss}Time-average toroidal magnetic field (TOP)
and the toroidal electromotive force (BOTTOM) shown in spatial distribution.}
\end{figure}

\begin{figure}[ht]
  \begin{center}
    \epsfig{file=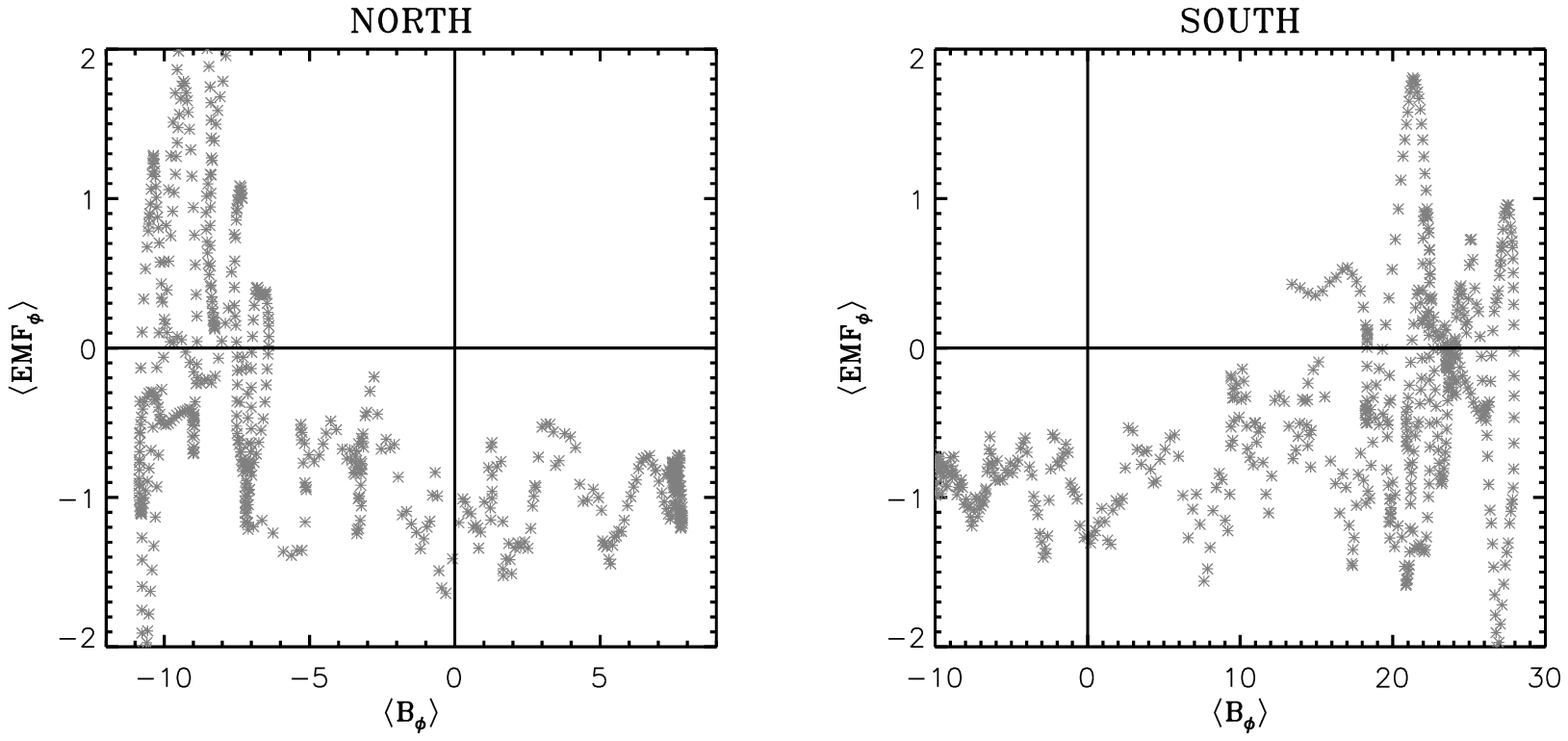, width=9.1cm}
    \epsfig{file=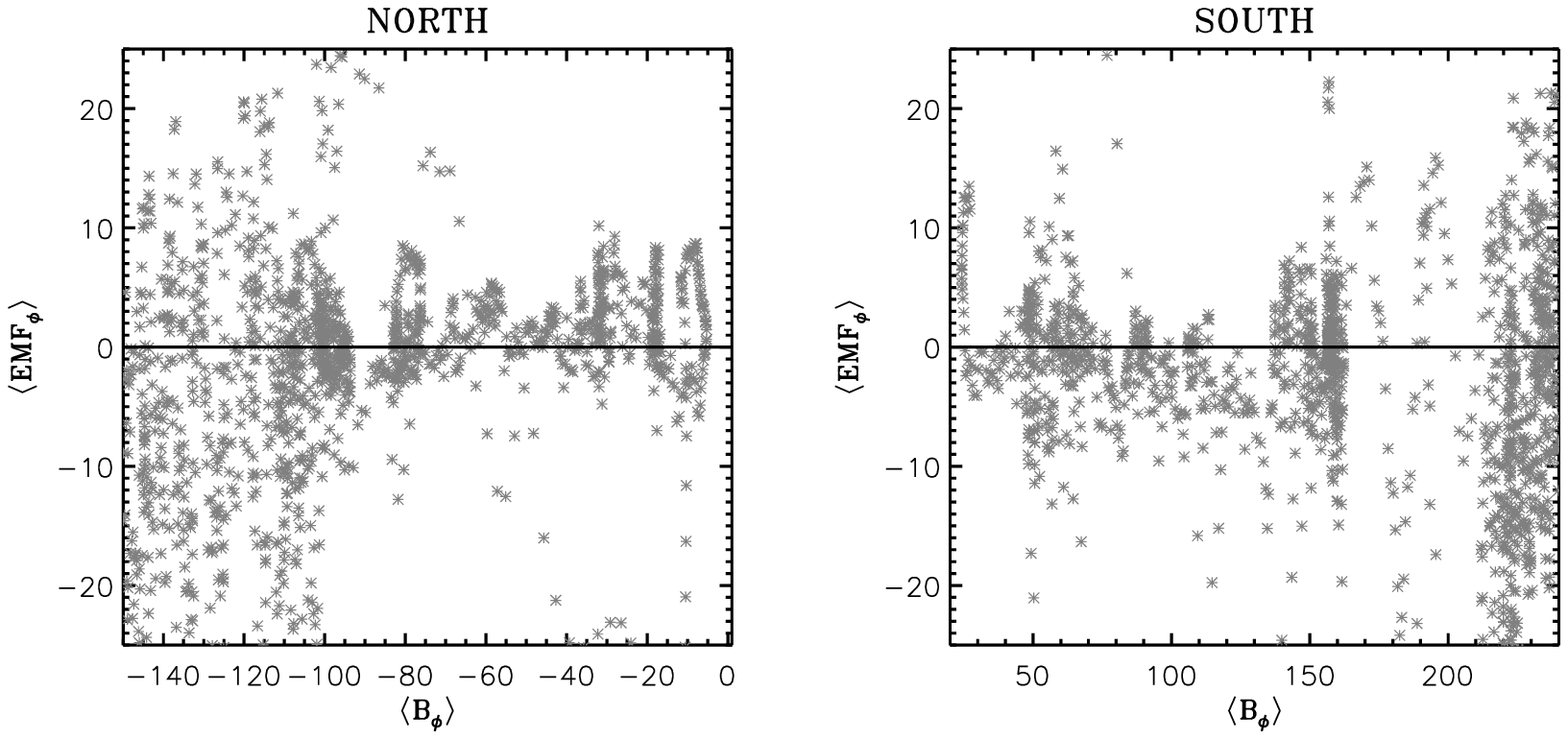, width=9.1cm}
    \epsfig{file=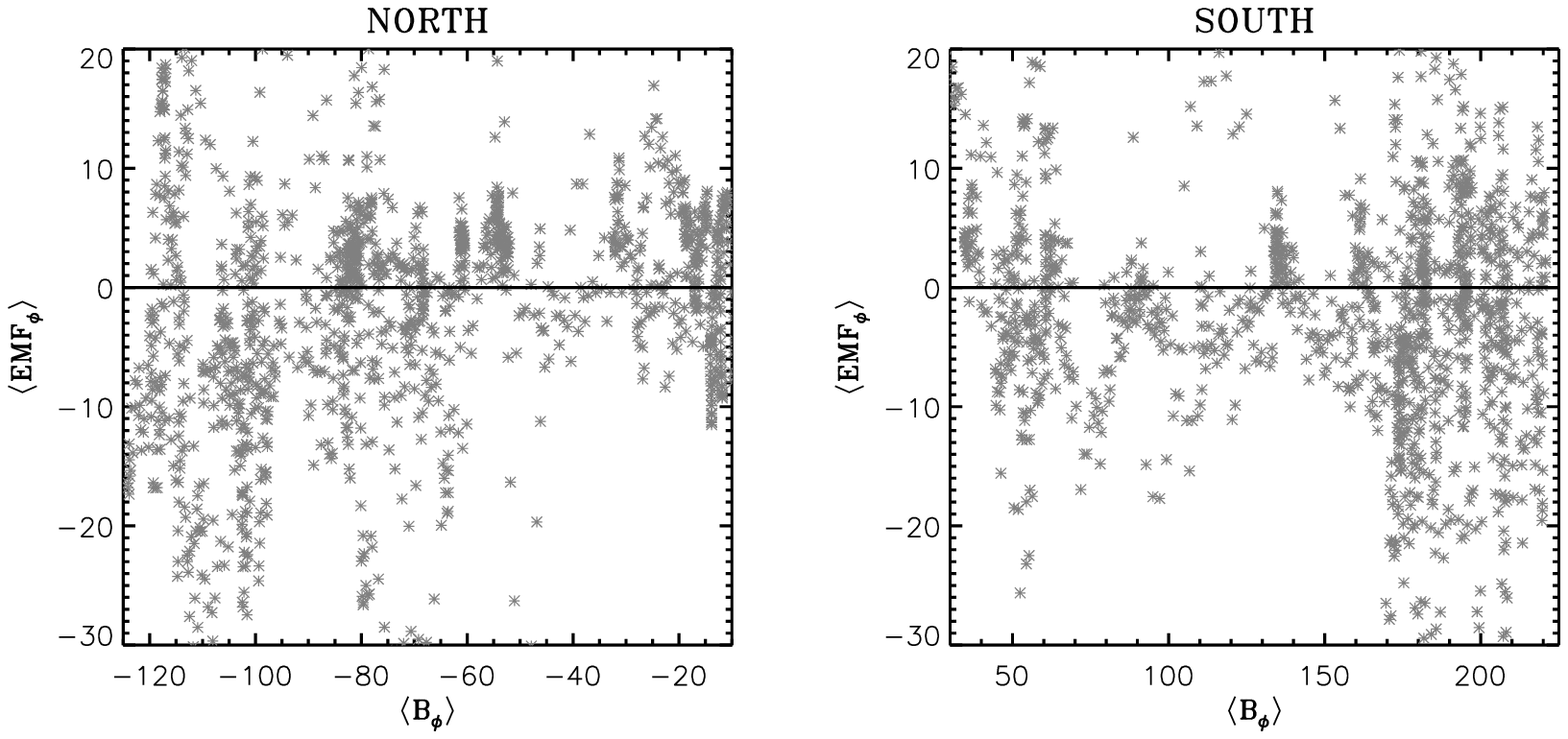, width=9.1cm}
  \end{center}
\caption{\label{dyn_temp2}Correlation of the average electromotive force versus 
the mean $B_\phi$, averaged for the northern and southern hemisphere
separately. The quantities were derived, from top to bottom, from the runs 
of Models Ia, V, and VI. The sign of the correlation gives an
indication for the sign of the dynamo $\alpha$-effect.}
\end{figure}

In order to evaluate the applicability of the $\alpha^{\rm dyn}$ approach,
we compute the average toroidal field $\langle B_\phi\rangle$
and compare it with the actual average electromotive force
${\cal E}_\phi=(\langle\vec{u}\rangle\times \langle\vec{B}\rangle)_\phi$
derived directly from the simulated vector fields. A correlation
between the two quantities may justify the $\alpha$-effect for
dynamo generation of magnetic fields. As a meaningful $\alpha^{\rm dyn}$ 
must change its sign at the equator, we plot the correlation
for both hemispheres of the disk separately; the result is given
in Figure~\ref{dyn_temp2}. As the temporal fluctuations are strong,
we plot the time-averages of $B_\phi$ and 
$(\vec{u}\rangle\times \langle\vec{B})_\phi$ in Figure~\ref{emf_bgauss}
only with their resulting sign. The quantities are also averaged 
in azimuthal direction and plotted in the $(r,z)$-plane. White areas
denote a negative sign; the dominance of negative averaged electromotive
forces indicates a positive $\alpha^{\rm dyn}$ in the northern 
hemisphere in accordance with Figure~\ref{dyn_temp2}.

The classical understanding of the Coriolis force giving
preference to left-handed helicity on the northern hemisphere
(whence positive $\alpha^{\rm dyn}$) appears not applicable for
the correlation plots of Models~II and III. The fluctuations
were strong, and no meaningful sign of $\alpha^{\rm dyn}$
can be derived. However, Models Ia, V, and VI show a mostly
negative averaged EMF for both hemispheres, while $\langle B_\phi\rangle$
changes its sign. An indication for positive $\alpha^{\rm dyn}$
is thus found from these simulations.

In fact, Models~V and VI are those which show saturated kinetic energies.
The other significant difference to the indefinite Models~II and III is 
the 10times higher diffusivity, $\eta=0.01$.
Notwithstanding, an indication for a negative $\alpha^{\rm dyn}$ 
was already found in simulations of a local box cut out of the 
disk by Brandenburg et al.\ (1995). The same sign is found by
Ziegler \& R\"udiger (2000b) from long runs of local simulations,
although a clear influence of resistivity on the evolution
was found there. Lowest magnetic Reynolds numbers are 
${\rm Re}_{\rm m}=1450$ for Model~V in this Paper, calculated 
with the velocity difference between inner and outer radial 
boundary, just as Ziegler \& R\"udiger did for the local box.

These considerations are somewhat limited, since a straight-forward 
regression line in Fig.~\ref{dyn_temp2} will have a 
significant offset from the origin of the graph; a vanishing
$\langle B_\phi\rangle$ does apparently not coincide with a
vanishing $(\langle\vec{u}\rangle\times \langle\vec{B}\rangle)_\phi$.
The effect of the other field components is considered small though;
while the average hemispherical $\langle B_\phi\rangle$ are of the
order of 100, the $\langle B_r\rangle$ is of order 10 and
$\langle B_z\rangle$ of order unity. A considerable contribution
is suspected from the current density through the $\eta_{ijk}$-term
in (\ref{emf}), and is appears to be quite natural that
an offset in the correlation graphs in Figure~\ref{dyn_temp2}
is found. For this reason, we did not plot such regression 
lines in the Figure.

Additionally, the sole consideration of the induction
equation in the ``classical'' dynamo theory leads to the omission
of the Lorentz force, which is not only essential for the onset
of the magneto-rotational instability but for the maintenance
of the turbulence and the transport of angular momentum as well.
It is thus obvious that the neglect of the back-reaction of
magnetic fields on the flow need not lead to representative
models of Keplerian disks.

\section{Conclusions}
Three-dimensional global simulations have shown, that the
magnetic shear-flow instability is a fast mechanism to 
generate a turbulent flow in a Keplerian disk. We presented
models with a computational domain covering the full azimuthal
range, accounting for accretion of matter. The Shakura-Sunyaev
parameter $\alpha_{\rm SS}$ increases rapidly during the
first revolutions after switching on the magnetic field and
reaches $10^{-2}$--$10^{-1}$ at this stage of the computations,
though it appears not yet fully saturated in several of the models. 
Maxwell stresses exceed Reynolds stresses almost entirely. The
latter undergo strong fluctuations and are partly negative.

Indications for a dynamo action in the disk are found,
the corresponding dynamo-$\alpha^{\rm dyn}$ tending to be negative north
of the equatorial plane and positive south of the equatorial
plane. The generated magnetic fields may maintain the turbulence
even if the external field ceases. The strong excitation of
low-order azimuthal modes in the magnetic field is another
promising fact for dynamo action with respect to the
Cowling theorem.

\subsection{The dynamo of Model~V}
For various reasons, a representative example for an accretion disk
dynamo is provided by Model~V (and nearly as well by the similar 
Model~VI with higher inflow velocity). The time series is sufficiently
long covering more than 18~revolution periods. The angular momentum
transport reaches a saturated level of high efficiency during the
last four orbital periods. The magnetic diffusivity is roughly
two orders of magnitude higher than the numerical diffusivity.
The striking dominance of one sign of the averaged EMF in 
Figure~\ref{emf_bgauss} indicates a physically evolved, relevant
state of the system.

The results connected with the outcomes of mean-field dynamo theory
include include the following facts: (i) a negative average toroidal 
magnetic field is found for the northern hemisphere, a positive on
the southern one. The averaged EMF is negative in both hemisphere
indicating a {\it positive\/} $\alpha^{\rm dyn}$-effect on the
northern side (negative on the southern side). (ii) The correlation
plot of average toroidal field and EMF gives the same picture.
(iii) {\it negative\/} kinetic and magnetic helicities on the
northern hemisphere (positive on the southern one) are also
consistent with a positive $\alpha^{\rm dyn}$.
(iv) The {\it dipolar structure\/} of the solution contrasts with the
results from mean-field dynamo models of disks which yield
quadrupolar solutions for positive $\alpha^{\rm dyn}$. We have
to add though that the disks of our Paper are {\it not thin\/}.
Mean-field simulations by Covas et al.\ (1999) show a transition from 
quadrupolar to dipolar fields when the opening angle of the disk
is enlarged, rather representing a torus than a disk.

The extension of the run-times of such simulations exceeding
the order of diffusion times promises further results about
dynamo action in accretion disks. The connection with mean-field
concepts are most interesting as well as the implications
for jet-launch models.

\begin{acknowledgements} 
The invaluable help by D.\ Elstner, Potsdam, in adapting the ZEUS-3D 
code for our purposes is greatly acknowledged.
We are grateful to the John v. Neumann-Institut for Computing at the
Forschungszentrum J\"ulich, Germany, for the opportunity to use 
the Cray T90 computer. 
R.A.\ thanks for the kind  support by the Deutsche Forschungsgemeinschaft.
\end{acknowledgements}

\end{document}